\documentclass[12pt,a4paper]{article}
\usepackage{amsfonts}
\usepackage{amsmath}
\usepackage{amsthm}
\usepackage{mathptmx}
\usepackage[dvipdfmx]{graphicx}
\def \td{\tilde{\delta}}
\def \lam{\lambda}
\def \Lam{\Lambda}
\def \del{\partial}
\def \sig{\sigma}
\def \eps{\varepsilon}

\def \s2{\sin^2 \! }

\def \c2{\cos^2 \! }

\def \St0/T{\sum_{k = 0} ^{\infty } D_{\del_{\mu } \del_{\mu_1 } \cdots \del_{\mu_k } f } ^{ t^0 / T} \sqrt{ - g } }

\def \seq{\overset{\mathrm{stationary}}{=}}
\def \sseq{\overset{\mathrm{static}}{=}}

\def \t{\tilde}
\def \d{{\mathrm d}}
\def \Bg{\, ^{\mathrm B} \! g}
\def \Fg{\, ^{\mathrm F} \! g}
\def \Bnabla{\, ^{\mathrm B} \! \nabla}
\def \D{\, ^{\rm EL} \! \mathcal{D}}
\def \2D{\, ^{\rm EL} _{ (2) } \! \mathcal{D}}
\def \3D{\, ^{\rm EL} _{ (3) } \! \mathcal{D}}

\def \bD{\, ^{\rm EL} \! \bar{ \mathcal{D} }}
\title{The maximum entropy principle of self-gravitating fluid system and field equations}
\author{Hikaru Yumisaki\thanks{e-mail: yumisaki722@gmail.com}\\
\ \\
{\it Gunma University, 3-39-22 Showa, Maebashi 371-8511, Japan}}

\begin{document}
\maketitle
\abstract{
We investigate the maximum entropy principle for general field theory, including a metric tensor $g_{ \mu \nu }$, a vector field $A_{ \mu }$, and a scalar field $\varphi$ as the fundamental fields, and find (i) imposing an ordinary constraint relation on $\eps$, the field equations, which is constructed of the Euler-Lagrange derivative of an arbitrary Lagrangian density, the stress tensor of a perfect fluid, and the electric current vector, in ordinary manner, are compatible with the maximum entropy principle, and (ii) varying also the constraint relation on $\eps$, the maximum entropy principle requires an extra scalar field $\eta$, which is introduced as the difference from the ordinary constraint relation on $\eps$. The field $\eta$ is also interpreted as the difference between two geometries, i.e., one is the geometry, defined by $g_{ \mu \nu }$, in which the thermodynamical relations are written in the ordinary and simplest form, and the other is the geometry, defined by 
$$\t g_{ \mu \nu } = e^{ 2 \eta } g_{ \mu \nu },$$
in which a force-free fluid flows along a geodesic orbit, instead the thermodynamical relations, the first law of thermodynamics and the Gibbs-Duhem relation, are modified. 

We also calculate the variation of the entropy $\delta S$ in the Lovelock theory, in which $\delta S$ is expressed as a linear combination of the variations of the Kodama energy and the size of the system. Finally, with the field $\eta$ introduced, we propose a set of field equations in the thermodynamical and kinematical geometries, which possesses appropriate scaling properties, and point out that in vacuum spacetime they resemble those of the dilaton gravity in the string and Einstein frame, respectively. 
}
\tableofcontents
%\newpage
\section{Introduction}

\subsection{Background and motivation}
One of the properties of gravity that leads to difficulties in studying thermodynamics of self-gravitating fluid is that it generates long range self-interactions. In statistical picture, long range self-interactions affect the additivity of the macroscopic quantities, and it becomes impossible to understand the phenomena of the whole system as a simple sum of small parts. Therefore, the traditional statistical procedure does not guarantee the validity of the simple combination of thermodynamics and a gravity theory. On the other hand, in purely thermodynamical picture, since the presence of long range self-interaction relates the thermodynamical phenomena and dynamics of fields, requiring the compatibility of the two theories, namely thermodynamics and field theory, may give us some perspectives for macroscopic long range self-interacting system. In this work, we focus on features of the final state of a system in thermodynamical and field theoretical pictures, namely equilibrium and stationary states, respectively, and investigate their compatibility.   

In thermodynamical picture, an essential feature of the final state of a system is stated by the concept of ``entropy''. The second law of thermodynamics states that, given an arbitrary isolated system $\mathcal{V}$, there exist a quantity $S$, which is called entropy, for each states such that the irreversibility of the system is represented by forbidding the entropy to decrease. Therefore, if there exists the final state of the system, called the equilibrium state, then the equilibrium state is the state of maximum entropy among all the {\it allowed states}, which is called ``the maximum entropy principle''. The range of the {\it allowed states} mentioned above depends on the give situation. 

On the other hand, including also kinematical picture, we believe that the final state of a system is static or more generally stationary. For example, in an ordinary field theory with a self-interact fluid system, if a state is not static nor stationary, namely dynamical, then the entropy of a system can increase without violating the energy and the other conservation laws by converting the kinetic energy into heat energy.\footnote{We have assumed the positivity of kinetic energy of the fields including the gravity, the conservation of total energy, and that the entropy is monotonically increasing function for heat energy. It is well known that the first two assumptions are delicate problems in gravitational field theories \cite{Gravitation}.} Therefore, it is obvious that the entropy of a static state is a maximum among those of all the states that satisfy the given field equations, so-called ``on-shell'' states, with the total energy and the other conserved charges fixed. However, it is a nontrivial problem whether the entropy of a static state is maximum among those of all the states that contains also the states that do {\it not} satisfy the field equations, namely ``off-shell'' states. 

Is it necessary that the entropy of a static solution of the given field equations is maximum among those of configurations including off-shell ones? Suppose that an infinitesimal variation increase the entropy of a static solution, then the second law of thermodynamics indicates that the ``static'' solution varies so as to increase its own entropy\footnote{In general, a force acting in a system resulting from the tendency to increase the total entropy is called an ``entropic force''.}, that is to say, the requirement of field theory conflicts with that of thermodynamics. One can interpret the situation as an instability of the static solution to thermal fluctuations. 

In this work, we adopt as the maximum entropy principle for field theory that {\it the entropy of an arbitrary static solution of the field equations should be maximum among those of all the configurations that satisfy given appropriate boundary conditions.} The aim of this study is to reveal the condition for field theories to be compatible with the maximum entropy principle. Indeed, one finds below that in such a field theory not only static solutions but also stationary ones satisfy the maximum entropy principle. 

Related calculations are found in literature. In \cite{Cocke1965}, it is stated that requiring to vanish the variation of the entropy among spherically symmetric solutions of the Einstein equation is consistent with the equation of hydrostatic equilibrium. In \cite{SWZ1981}, for a spherically symmetric and static radiation fluid system in general relativity, it is shown that the initial value constraint equations (equivalent to the time-time component of the Einstein equation) and requiring for the total entropy to be maximum with the total mass fixed derive the Tolman-Oppenheimer-Volkoff (TOV) equation of hydrostatic equilibrium. Generalizations have been done, i.e., for arbitrary equation of state of fluids \cite{Gao2011}, non-spherically symmetric system \cite{FangGao2014}, and attempts to apply other field theories, namely, the Lovelock theory \cite{CaoZeng2013}, the $f(R)$ theory \cite{FGJ2016}, Einstein-Maxwell theory \cite{FG2015}, and generally covariant purely gravitational theories \cite{JFG2021}. However, since these calculations are based on variations restricted by some constraint conditions such as spherical or time translational symmetries, Tolman's law, or the time-time component of the gravitational field equation, calculations for general field theories and based on general variations have not yet done. 

In this paper, in order to keep the generality, we do not use any assumptions about the field equations, except for that the thermodynamical quantities, including the entropy, are functions of a symmetric second-order tensor field $g^{ \mu \nu }$, a vector field $A_{ \mu }$, some scalar fields, and their arbitrarily high order derivatives. 

This paper is organized as follows. In section \ref{Therm and second law}, using thermodynamical relations, we calculate the first order variation of the entropy. In section \ref{ELoperator}, the Euler-Lagrange operator is generalized to deal with functions that contain arbitrarily high order derivatives of fields. In section \ref{Volume term}, we consider the volume term of the variation of the entropy, and derive the conditions for the stationary states to be equilibrium. Redefining the metric tensor and extensive quantities in order for them to have physically reasonable features, for example force-free fluids flow along geodesic orbits, we introduce the kinematical geometry, which is described by the redefined metric tensor $\t g_{ \mu \nu }$. In this frame, we obtain the field equations (\ref{ELeqGtilde}) - (\ref{ELeqphitilde}) for arbitrary Lagrangian that are compatible with the maximum entropy principle. In section \ref{Surface term}, we consider the surface term of the variation of the entropy. Each surface term is interpreted as the product of a ``chemical potential'' and the variation of the conjugate quantity. As an example, we explicitly calculate the variation of the entropy of a spherically symmetric system in the Lovelock theory and show that the variation of the entropy is expressed as a linear combination of those of the Kodama energy and the size of the system. In section \ref{Constraint Lagrangian}, a Lagrangian which has thermodynamically reasonable properties is presented. In section \ref{Conclusion}, we summarize the conclusions, and briefly discuss their applications to other topics, such as cosmology, thermodynamical aspects of black hole, and so on. 

\subsection{Conventions and notations}
In this paper, we use symbols in the following conventions. The dimension of spacetime is denoted by $d$. The signature of Lorentzian metric tensor obeys the mostly plus convention $( - , + , \cdots , + )$. Greek indexes $\alpha, \beta, \cdots$ span the full dimension of spacetime, namely $\{ 0 , 1 , \cdots , d - 1 \}$. The index $0$ often represents a timelike coordinate, whereas Latin indexes $a, b, \cdots$ do spacelike ones, so $a, b, \cdots \in \{ 1 , \cdots , d \}$. 

The Levi-Civita symbol $\epsilon _{ \mu_1 \cdots \mu_d } $ is defined by 
\begin{eqnarray}
\epsilon _{ \mu_1 \cdots \mu_d }  =  
  \begin{cases}
  + 1 \ \ \ \ \ \  {\rm if \ } ( \mu_1 , \cdots , \mu_d ) {\rm \ is\ an\ even\ permutation\ of\ } ( 0 , 1 , \cdots , d -1 ) , & \\
  - 1 \ \ \ \ \ \  {\rm if \ } ( \mu_1 , \cdots , \mu_d ) {\rm \ is\ an\ odd\ permutation\ of\ } ( 0 , 1 , \cdots , d -1 ) , & \\
  \ \ \ 0 \ \ \ \ \ \  {\rm otherwise, } & 
  \end{cases}
\end{eqnarray}
in any coordinates, which is a $( 0 , d )$-type covariant tensor density of weight $- 1$. 

We use the symbols which have $n$ subscripts for the bases of $( d - n )$-forms defined by 
\begin{eqnarray}
( \d ^{ d - n } x ) _{ \mu_1 \cdots \mu_n }  :=  \frac{ 1 }{ ( d - n ) ! }  \epsilon _{ \mu_1 \cdots \mu_n \nu_1 \cdots \nu_{ d - n } } \d x ^{ \nu_1 } \wedge \cdots \wedge \d x ^{ \nu_{ d - 1 } } , 
\end{eqnarray}
so that a $(n , 0)$-type covariant tensor density $\sqrt{ - g } X^{ \mu_1 \cdots \mu_n }$ is naturally associated with a $(d - n)$-form $\bold X$, i.e., 
\begin{eqnarray}
\bold X  =  \frac{ 1 }{ n ! } ( \d ^{ d - n } x ) _{ \mu_1 \cdots \mu_n } \sqrt{ - g } X^{ \mu_1 \cdots \mu_n } . 
\end{eqnarray}

A set of fields which are taken for the fundamental ones are denoted by $f$. For example, when we think of $g^{ \mu \nu }$, $A_{ \mu }$, and $\varphi$ as fundamental fields, then $f = \{ g^{ \mu \nu } , A_{ \mu } , \varphi \}$.

\section{Thermodynamical relations and the second law} \label{Therm and second law}

Consider a $( d - 1 )$-dimensional region $\mathcal{V}$ of a fluid system in $d$-dimensional spacetime $\mathcal{M}$. The $d$-velocity of the fluid is denoted by $u^{ \mu }$. The entropy $S$ of the region $\mathcal{V}$ is defied as a real number such that we can represent the irreversibility of $\mathcal{V}$ by asserting that the quantity $S$ never decrease, which is called the second law of thermodynamics. We assume that the entropy $S$ always exists and is written in the form 
\begin{eqnarray}
S  =  \int _{ \mathcal{V} } ( \d ^{ d - 1 } x ) _{ \mu } \sqrt{ - g } u ^{ \mu } s( \eps , n ) , 
\end{eqnarray}
for an arbitrary region $\mathcal{V}$. We employ a second-order symmetric Lorentzian tensor $g_{ \mu \nu }$ and its determinant $g$. The type of the fluid determines the form of the scalar function $s(\eps , n)$ which depends on two scalar quantities $\eps $ and $n$. The generalization to the case that there exist more than one $n$ is straightforward. Since we do not specify the type of fluid, $s( \eps , n )$ is an arbitrary function. The quantity $\sqrt{ - g } u^{ \mu } s$ is called the entropy current density.

The $d$-velocity of the fluid $u^{\mu } $ is normalized by 
\begin{eqnarray}
g_{\mu \nu } u^{\mu } u^{ \nu }  =  - 1 . \label{normu}
\end{eqnarray}

In this paper, we often use a coordinate, called a ``co-moving coordinate'', in which the $d$-velocity $u^{ \mu }$ is parallel to the time axis, namely $u^{ \mu } \del_{ \mu } = \sqrt{ - g^{ 00 } } \del_0$, and the hypersurface defied by $x^0 = 0$ contains the region $\mathcal{V}$, i.e., $\{ x^1 , \cdots , x^{ d - 1 } \}$ is a coordinate of $\mathcal{V}$.

Supposing that the conservation law of the entropy current density 
\begin{eqnarray}
\del _{\mu } \Big( \sqrt{ - g } u^{\mu } s \Big)  =  0 \label{consvS}
\end{eqnarray}
is satisfied, and the boundary of the system $\del \mathcal{V}$ is fixed, then the entropy $S$ does not depend on the choice of the region $\mathcal{V}$. We also assume the conservation law of $\sqrt{ - g } u^{ \mu } n $, namely, 
\begin{eqnarray}
\del_{ \mu } \Big( \sqrt{ - g } u^{ \mu } n \Big)  =  0 , \label{consvN}
\end{eqnarray}
so that the quantity 
\begin{eqnarray}
N  =  \int _{ \mathcal{V} } \big( \d ^{ d - 1 } x \big) _{ \mu }  \sqrt{ - g } u^{ \mu } n  
\end{eqnarray}
is a conserved charge. In section \ref{consistency}, we check whether Eqs. (\ref{consvS}) and (\ref{consvN}) hold, or not.  

Define the ``intensive'' quantities $T$, $\mu$, and $p$ by 
\begin{eqnarray}
\frac{ 1 }{ T }  &:=&  \frac{ \del s }{ \del \eps }  \label{defT} \\
\frac{ \mu }{ T }  &:=&  - \frac{ \del s }{ \del n }  \label{defMu} \\
p  &:=&  T s + \mu n - \eps . \label{defp}
\end{eqnarray}
These relations are often represented in the form 
\begin{eqnarray}
\delta \eps = T \delta s + \mu \delta n \label{FL} \\
Ts + \mu n = \eps + p \label{GDR} .  
\end{eqnarray}
The relation (\ref{FL}) is a local representation of the first law of thermodynamics, and (\ref{GDR}) is the Gibbs-Duhem relation. The relation (\ref{GDR}) comes from the extensive properties of the extensive quantities for sufficiently small region (See appendix \ref{Homogeneity} for details). Comparing Eq. (\ref{FL}) and the variation of Eq. (\ref{GDR}), one finds the constraint relation among the variation of the intensive quantities: 
\begin{eqnarray}
\delta p  =  s \delta T  +  n \delta \mu . \label{relationPTMu}
\end{eqnarray}

Here, we emphasize that although usually $g_{ \mu \nu }$, $\eps$, $n$, $T$, $\mu$, and $p$ are interpreted as the metric tensor, energy density, particle number density, temperature, chemical potential, and pressure, respectively, we have not said about the physical meanings of them yet, except for the volume element $\sqrt{ - g }$, which is introduced respecting the additivity of the entropy. In this work, we call the geometry defined by $g_{ \mu \nu }$ the ``thermodynamical geometry''. In section \ref{redef}, we will discuss whether the interpretation of these quantities are appropriate even if the kinematical features are taken into account, or not.

Let us calculate the variation of the entropy 
\begin{eqnarray}
\delta S  =  \int _{\mathcal{V}} ( \d ^{d -1} x ) _{ \mu }  \delta \Big( \sqrt{ - g } u^{ \mu } s \Big) . 
\end{eqnarray}
Using Eq. (\ref{FL}), (\ref{GDR}), and the Leibniz rule of $\delta$, the integrand of $\delta S$ is arranged as follows: 
\begin{eqnarray}
\delta \Big( \sqrt{- g} u^{\mu } s \Big) &=& \sqrt{- g} u^{\mu } \delta s + s \delta \Big( \sqrt{- g} u^{\mu } \Big) \\
	&=& \sqrt{- g} \frac{u^{\mu } }{T} (\delta \eps - \mu \delta n)  +  \frac{\eps + p - \mu n }{T} \delta \Big( \sqrt{ - g } u^{\mu } \Big) \\
	&=&  \frac{1}{T} \bigg[ \sqrt{ - g } u^{\mu } \delta \eps + (\eps + p) \delta \Big( \sqrt{- g} u^{\mu } \Big) - \mu \delta \Big( \sqrt{- g} u^{\mu } n \Big)  \bigg] \\
	&=&  \frac{1}{T} \delta \Big( \sqrt{ - g } u^{\mu } \eps  \Big) + \frac{p}{T} \delta \Big( \sqrt{ - g } u^{\mu } \Big) - \frac{\mu }{T} \Big( \sqrt{ - g } u^{\mu } n \Big) . 
\end{eqnarray}
Decomposing the variation by the Leibniz rule,  
\begin{eqnarray}
\delta \Big( \sqrt{- g} u^{\mu } s \Big)  =  - \frac{\mu }{T} \delta \Big( \sqrt{ - g } u^{\mu } n \Big)  +  \frac{1}{T} \bigg[  u^{ \mu } \delta \Big( \sqrt{ - g } \eps \Big) +  \sqrt{ - g }  \eps \delta u^{ \mu }  +  p \Big( \sqrt{ - g } \delta u^{ \mu }  +   u^{ \mu }  \delta \sqrt{ - g }  \Big)  \bigg] . \label{varyS1}
\end{eqnarray}
The relations among the variations of $u^{ \mu }$, $\sqrt{ - g }$, and $g^{ \mu \nu }$ are required. Varying the normalization condition (\ref{normu}), 
\begin{eqnarray}
u_{\mu } \delta u^{\mu }  &=&  - \frac{1}{2} u^{\mu } u^{\nu } \delta g_{\mu \nu }  \\
	&=&  \frac{1}{2} u_{\mu } u_{\nu } \delta g^{\mu \nu } , 
\end{eqnarray}
where we use the relation $g^{ \mu \rho } g_{ \rho \nu }  =  \delta ^{ \mu } _{ \nu }$. 
Therefore, the variations of $u^{\mu } $ can be written in the form 
\begin{eqnarray}
\delta u^{\mu }  =  - \Big( \frac{1}{2} u_{\rho } u_{\sig } \delta g^{\rho \sig } \Big) u^{\mu }  +  \big(  u^{ \mu } u_{ \nu }  +  \delta ^{\mu }  _{\nu }  \big) \delta u^{\nu } . \label{varyU0}
\end{eqnarray}
Notice that $ - u^{ \mu } u_{ \nu } $ is the projection operator on the direction parallel to $u^{ \mu }$, and $u^{ \mu } u_{ \nu }  +  \delta ^{ \mu } _{ \nu } $ is that of the $( d - 1 )$-dimensional hypersurface orthogonal to $u^{ \mu }$. The first term of the right hand side of Eq. (\ref{varyU0}) represents the variation through the change of the normalization condition, whereas the second term does that of the direction of $d$-velocity $u^{\mu } $. As we are interested in the variation with the direction of $d$-velocity $u^{ \mu }$ fixed, the second term of Eq. (\ref{varyU0}) is ignored, namely 
\begin{eqnarray}
\delta u^{\mu }  =  - \Big( \frac{1}{2} u_{\rho } u_{\sig } \delta g^{\rho \sig } \Big) u^{\mu } . \label{varyU}
\end{eqnarray}
The variation of $u_{ \mu }$ is 
\begin{eqnarray}
\delta u_{ \mu }  &=& \delta \big( g_{ \mu \nu } u^{ \nu } \big)  \\
	&=&  - u_{ \rho } \Big( \frac{ 1 }{ 2 } u_{ \mu } u_{ \sig }  +  g_{ \mu \sig } \Big)  \delta g^{ \rho \sig }  . 
\end{eqnarray}
The variation of the projection operator $- u_{ \mu } u^{ \nu }$ is 
\begin{eqnarray}
\delta \big( - u_{ \mu } u^{ \nu } \big)  =   u^{ \nu } u_{\rho } \big( u_{ \mu } u_{ \sig }  +  g_{ \mu \sig } \big)  \delta g^{ \rho \sig } .  \label{varyUU}
\end{eqnarray}
Substituting Eq. (\ref{varyU}) and the formula \cite{Gravitation} 
\begin{eqnarray}
\delta \sqrt{ - g }  =  - \frac{1}{2} \sqrt{ - g } g_{\mu \nu } \delta g^{\mu \nu } , \label{varySQg}
\end{eqnarray}
into Eq. (\ref{varyS1}), one obtains the following variational equation:  
\begin{eqnarray}
&& \delta \Big( \sqrt{- g} u^{\mu } s \Big)   \nonumber \\
	&=&  - \lam \delta \Big( \sqrt{ - g } u^{\mu } n \Big)  +  \frac{u^{\mu } }{T}  \bigg\{  \delta \Big( \sqrt{ - g } \eps \Big) - \frac{1}{2} \sqrt{ - g } \Big[  \eps  u_{\rho } u_{\sig } + p \big( u_{ \rho } u_{ \sig }  +  g_{\rho \sig }  \big)  \Big] \delta g^{\rho \sig }  \bigg\}  , \nonumber \\
\end{eqnarray}
where the quantity $\lam$ is introduced by 
\begin{eqnarray}
\lam :=  \frac{ \mu }{ T } . 
\end{eqnarray}

\section{The Euler-Lagrange operator} \label{ELoperator}

In this work, we deal with the variations of quantities which contain arbitrarily high order derivatives of fields, so that the higher-order Euler-Lagrange procedures often appear in the following calculations. In this section, we define the Euler-Lagrange operator $\D _f$ and present some formulae. 

\subsection{Definition}
Consider the variation of a quantity $A$ that depends on arbitrarily high order derivatives of fundamental fields $f$. Using the Leibniz's rule of the partial derivative $\del_{ \mu }$ and the commutative relation $\delta \del_{ \mu } =  \del_{ \mu } \delta $, the first-order variation of a function $A$ is expanded in the following form: 
\begin{eqnarray}
\delta A   &=&   \sum_{ j = 0 } ^{ \infty } \frac{ \del A }{ \del ( \del_{ \mu_1 } \cdots \del_{ \mu_j } f ) }  \delta \del_{ \mu_1 } \cdots \del_{ \mu_j } f  \label{varyexpansion} \\
	&=&   \sum _{ j = 0 } ^{ \infty } \Bigg[  \del_{ \mu_1 } \bigg( \frac{ \del A }{ \del ( \del_{ \mu_1 } \cdots \del_{ \mu_j } f ) } \delta \del_{ \mu_2 } \cdots \del_{ \mu_j } f   \bigg)  -  \del_{ \mu_1 } \frac{ \del A }{ \del ( \del_{ \mu_1 } \cdots \del_{ \mu_j } f ) } \delta \del_{ \mu_2 } \cdots \del_{ \mu_j } f  \Bigg]  \nonumber \\
	&=&   \sum _{ j = 0 } ^{ \infty } \Bigg[  \del_{ \mu_1 } \bigg( \frac{ \del A }{ \del ( \del_{ \mu_1 } \cdots \del_{ \mu_j } f ) } \delta \del_{ \mu_2 } \cdots \del_{ \mu_j } f   \bigg)  -  \del_{ \mu_2 } \bigg( \del_{ \mu_1 } \frac{ \del A }{ \del ( \del_{ \mu_1 } \cdots \del_{ \mu_j } f ) } \delta \del_{ \mu_3 } \cdots \del_{ \mu_j } f   \bigg) \nonumber \\
	&&   +  \del_{ \mu_2 } \del_{ \mu_1 } \frac{ \del A }{ \del ( \del_{ \mu_1 } \cdots \del_{ \mu_j } f ) } \delta \del_{ \mu_3 } \cdots \del_{ \mu_j } f  \Bigg]  \\
	&=&  \cdots \nonumber \\
	&=&  \sum _{ j = 0 } ^{ \infty } \Bigg[  \sum _{ i = 0 } ^{ j - 1 } \del_{ \mu _{ i + 1 } } \bigg( ( - 1 )^{ i } \del_{ \mu _{ i } } \cdots \del_{ \mu _1 } \frac{ \del A }{ \del ( \del_{ \mu_1 } \cdots \del_{ \mu _ { i + 1 } } \cdots \del_{ \mu_j } f ) }  \delta \del_{ \mu_{ i + 2 } } \cdots \del_{ \mu_j } f  \bigg)  \nonumber \\
	&&  +  ( - 1 ) ^j \del_{ \mu_j } \cdots \del_{ \mu_1 }  \frac{ \del A }{ \del ( \del_{ \mu_1 } \cdots \del_{ \mu_j } f ) }  \delta f  \Bigg]  . 
\end{eqnarray}

Notice that in variational calculation, the quantities $\del_{ \mu_1 } \cdots \del_{ \mu_j } f$ are regarded as independent quantities each other, after that those quantities are identified. Therefore, there are ambiguities in the partial derivatives in terms of $\del_{ \mu_1 } \cdots \del_{ \mu_j } f$. In this work, those ambiguities are removed so that all the (anti-)symmetries in the arguments are inherited by the partial derivatives, for example
\begin{eqnarray}
\frac{ \del A }{ \del ( \del_{ \mu_1 } \cdots \del_{ \mu_j } f ) }  =  \frac{ \del A }{ \del ( \del_{ \mu_{ P( 1 ) } } \cdots \del_{ \mu_{ P( j ) } } f ) } , 
\end{eqnarray}
where $P( i )$ is a permutation of $i$.

Introducing a subscript $k$ by 
\begin{eqnarray}
j  =  i + k + 1 , 
\end{eqnarray}
the variation of $A$ is written in the following form 
\begin{eqnarray}
\delta A   &=&  \sum _{ k = 0 } ^{ \infty }  \sum _{ i = 0 } ^{ \infty } \del_{ \mu _{ i + 1 } } \bigg( ( - 1 )^{ i } \del_{ \mu _{ i } } \cdots \del_{ \mu _1 } \frac{ \del A }{ \del ( \del_{ \mu_1 } \cdots \del_{ \mu _ { i + 1 } } \cdots \del_{ \mu_{ i + k + 1 } } f ) }  \delta \del_{ \mu_{ i + 2 } } \cdots \del_{ \mu_{ i + k + 1 } } f  \bigg)  \nonumber \\
	&&  +  \sum_{ j = 0 } ^{ \infty } ( - 1 ) ^j \del_{ \mu_j } \cdots \del_{ \mu_1 }  \frac{ \del A }{ \del ( \del_{ \mu_1 } \cdots \del_{ \mu_j } f ) }  \delta f  . 
\end{eqnarray}
Defining the Euler-Lagrange operator $\D _f $ by 
\begin{eqnarray}
\D _f A   :=   \sum_{ j = 0 } ^{ \infty } ( - 1 ) ^j  \del_{ \mu_j } \cdots \del_{ \mu_1 } \frac{ \del A }{ \del ( \del_{ \mu_1 } \cdots \del_{ \mu_j } f ) } , \label{ELderivative}
\end{eqnarray}
the variation of $A$ can be written as 
\begin{eqnarray}
\delta A   =   \D _f A \cdot \delta f  +  \del_{ \mu } \bigg( \sum_{ k = 0 } ^{ \infty } \D _{ \del_{ \mu } \del_{ \mu_1 } \cdots \del_{ \mu_k } f } A  \cdot \delta \del_{ \mu_1 } \cdots \del_{ \mu_k } f  \bigg) . \label{ELexpansion}
\end{eqnarray}
We call $\D _f A$ the Euler-Lagrange ``derivative'' \footnote{The Euler-Lagrange derivative $\D _f $ does not satisfy the Leibniz's rule.} of $A$ with respect to $f$.

The commutative relation $\delta$ and $\del_{ \mu }$ derives some formulae. From Eq. (\ref{varyexpansion}) and (\ref{ELexpansion}), the variation of $\del_{ \mu } A$ and the divergence of $\delta A^{ \mu }$ are written in the following forms respectively: 
\begin{eqnarray}
\delta \del_{ \mu } A^{ \mu }  =  \D _f \del_{ \mu } A^{ \mu }  \cdot  \delta f  +  \del_{ \rho } \bigg( \sum _{ k = 0 } ^{ \infty } \D _{ \del_{ \rho } \del_{ \mu_1 } \cdots \del_{ \mu_k } f } \del_{ \mu } A^{ \mu }  \cdot \delta \del_{ \mu_1 } \cdots \del_{ \mu_k } f  \bigg) , 
\end{eqnarray}
and 
\begin{eqnarray}
\delta \del_{ \mu } A^{ \mu }  =  \del_{ \rho } \delta A^{ \rho }  =  \del_{ \rho } \bigg( \sum _{ k = 0 } ^{ \infty } \frac{ \del A^{ \rho } }{ \del ( \del_{ \mu_1 } \cdots \del_{ \mu_k } f ) }  \delta \del_{ \mu_1 } \cdots \del_{ \mu_k } f  \bigg) . 
\end{eqnarray}
Comparing the two representations, one obtains following two formulae: 
\begin{eqnarray}
\D _f \del_{ \mu } A^{ \mu }  &=&  0  , \\
\D _{ \del_{ \rho } \del_{ \mu_1 } \cdots \del_{ \mu_k } f } \del_{ \mu } A^{ \mu }  &=&  \frac{ \del A^{ \rho } }{ \del ( \del_{ \mu_1 } \cdots \del_{ \mu_k } f ) }  .  
\end{eqnarray}

When the variable $f$ is a function of $h$ that does not depend on the derivatives of $h$, namely $f = f(h)$, 
\begin{eqnarray}
\delta A &=& \D_f  A \cdot \delta f + \del_{\mu }  \bigg( \sum_{k = 0} ^{\infty } \D_{\del_{\mu } \del_{\mu_1 } \cdots \del_{\mu_k } f } A \cdot \delta \del_{\mu_1 } \cdots \del_{\mu_k } f \bigg) \\
	&=& \D_f A \cdot f_{, h } \delta h + \del_{\mu }  \bigg[  \sum_{k = 0} ^{\infty } \D_{\del_{\mu } \del_{\mu_1 } \cdots \del_{\mu_k } f }  A \cdot \del_{\mu_1 } \cdots \del_{\mu_k } \big( f_{, h } \delta h  \big)   \bigg] \\
	&=& \D_f  A \cdot f_{ , h } \delta h + \del_{\mu }  \bigg[  \sum_{ k = 0 } ^{\infty } \sum_{ j = 0 } ^k 
	\bigg(
    \begin{array}{c}
      k \\
      j
    \end{array}
	\bigg)  
  \D_{\del_{\mu } \del_{\mu_1 } \cdots \del_{\mu_k } f }  A \cdot  \del_{\mu_{j+1} } \cdots \del_{\mu_k } f_{ , h } \cdot \delta \del_{\mu_1 } \cdots \del_{\mu_j } h  \bigg] . \nonumber \\ \label{varyAh}
\end{eqnarray}
Comparing Eq. (\ref{varyAh}) with 
\begin{eqnarray}
\delta A = \D_h  A \cdot \delta h + \del_{\mu }  \bigg( \sum_{j = 0} ^{\infty } \D_{\del_{\mu } \del_{\mu_1 } \cdots \del_{\mu_j } h }  A \cdot \delta \del_{\mu_1 } \cdots \del_{\mu_j } h \bigg) , 
\end{eqnarray}
one obtains the transformation formulae for change of variables: 
\begin{eqnarray}
\D_h  A &=& f_{ , h } \D_f  A \\
\D_{\del_{\mu_1 } \cdots \del_{\mu_i } h }  A &=& \sum_{ k = i } ^{\infty } 
	\bigg(
		\begin{array}{c}
		k \\
		i
		\end{array}
	\bigg)  
	\big( \del_{\mu_{i+1} } \cdots \del_{\mu_k } f_{ , h } \big) \D_{ \del_{\mu_1 } \cdots \del_{\mu_k } f }  A \ \ \ \ \ \ ( i \ge 1 ) . 
\end{eqnarray}

More generally, when the variable $f$ is a function of $h$, $\del_{ \mu } h$, and their higher derivatives, namely $f  =  f ( h , \del_{ \mu } h , \cdots )$, the variations of $f$, $h$, and their derivatives are related by 
\begin{eqnarray}
\delta f  =  \frac{ \del f }{ \del h } \delta h  +  \frac{ \del f }{ \del ( \del_{ \mu } h ) } \delta \del_{ \mu } h  +  \cdots . 
\end{eqnarray}
The variation of a quantity $A$ is rearranged as 
\begin{eqnarray}
\delta A  &=&  A_f \delta f  +  \del_{ \mu } \bigg( \sum _{ k = 0 } ^{ \infty }  A_{ \del _{ \mu } \del_{ \mu_1 } \cdots \del_{ \mu_k } f } \cdot \del_{ \mu_1 } \cdots \del_{ \mu_k } f  \bigg)  \\
	&=&  \bigg[ A_f \frac{ \del f }{ \del h }  -  \del_{ \mu } \bigg( A_f \frac{ \del f }{ \del ( \del_{ \mu } h ) } \bigg) +  \cdots  \bigg] \delta h  \nonumber \\
	&&  +  \del_{ \mu } \bigg[ \sum _{ k = 0 } ^{ \infty }  A_{ \del _{ \mu } \del_{ \mu_1 } \cdots \del_{ \mu_k } f } \cdot \del_{ \mu_1 } \cdots \del_{ \mu_k } f  +  A_f \frac{ \del f }{ \del ( \del_{ \mu } h ) } \delta h  +  \cdots   \bigg]  \\
	&=:&  B_h \delta h  +  \del_{ \mu } \bigg( \sum _{ k = 0 } ^{ \infty }  B_{ \del _{ \mu } \del_{ \mu_1 } \cdots \del_{ \mu_k } h } \cdot \del_{ \mu_1 } \cdots \del_{ \mu_k } h  \bigg), 
\end{eqnarray}
where $A_f$, $A_{ \del_{ \mu } f }$, $\cdots$ are arbitrary coefficients. In general, two quantities 
\begin{eqnarray}
\del_{ \mu } \bigg( \sum _{ k = 0 } ^{ \infty }  A_{ \del _{ \mu } \del_{ \mu_1 } \cdots \del_{ \mu_k } f } \cdot \del_{ \mu_1 } \cdots \del_{ \mu_k } f  \bigg)
\end{eqnarray}
and 
\begin{eqnarray}
\del_{ \mu } \bigg( \sum _{ k = 0 } ^{ \infty }  B_{ \del _{ \mu } \del_{ \mu_1 } \cdots \del_{ \mu_k } h } \cdot \del_{ \mu_1 } \cdots \del_{ \mu_k } h  \bigg)
\end{eqnarray}
are different. However, when the coefficient $A_f$ vanishes, then $B_h$ also vanishes and the above two coincide: 
\begin{eqnarray}
\del_{ \mu } \bigg( \sum _{ k = 0 } ^{ \infty }  A_{ \del _{ \mu } \del_{ \mu_1 } \cdots \del_{ \mu_k } f } \cdot \del_{ \mu_1 } \cdots \del_{ \mu_k } f  \bigg)  =  \del_{ \mu } \bigg( \sum _{ k = 0 } ^{ \infty }  B_{ \del _{ \mu } \del_{ \mu_1 } \cdots \del_{ \mu_k } h } \cdot \del_{ \mu_1 } \cdots \del_{ \mu_k } h  \bigg) . 
\end{eqnarray}

\subsection{The variation of $g^{ \mu \nu } $ and $u^{ \mu } $}
Since the normalization condition of $u^{ \mu }$ depends on $g^{ \mu \nu }$, the variation of quantities that depend on $u^{ \mu }$ and $g^{ \mu \nu }$ should be calculated carefully. 

Using the relation (\ref{ELexpansion}), the variation of $A = A [ g^{ \mu \nu } , u^{ \mu } ] $ is arranged as 
\begin{eqnarray}
\delta A  &=&  \D _{ g^{ \mu \nu } } A  \cdot \delta g^{ \mu \nu }  +  \D _{ u^{ \mu } } A  \cdot \delta u^{ \mu }  +  \del _{ \rho } \bigg( \sum_{ k = 0 } ^{ \infty } \D _{ \del_{ \rho } \del_{ \mu_1 } \cdots \del_{ \mu_k } f } A  \cdot  \delta \del_{ \mu_1 } \cdots \del_{ \mu_k } f \bigg)  \\
	&=&  \Big( \D _{ g^{ \mu \nu } } A  -  \frac{1}{2} u_{ \mu } u_{ \nu } u^{ \rho } \D _{ u^{ \rho } } A \Big) \cdot \delta g^{ \mu \nu }  +  \del _{ \rho } \bigg( \sum_{ k = 0 } ^{ \infty } \D _{ \del_{ \rho } \del_{ \mu_1 } \cdots \del_{ \mu_k } f } A  \cdot  \delta \del_{ \mu_1 } \cdots \del_{ \mu_k } f \bigg) .  \nonumber \\
\end{eqnarray}
Defining the new operator $\bD _{ g^{ \mu \nu } }$ by 
\begin{eqnarray}
\bD _{ g^{ \mu \nu } }  :=  \D _{ g^{ \mu \nu } }   -  \frac{1}{2} u_{ \mu } u_{ \nu } u^{ \rho } \D _{ u^{ \rho } } , \label{defDbar}
\end{eqnarray}
the variation of $A$ is written in the form 
\begin{eqnarray}
\delta A  &=&  \bD _{ g^{ \mu \nu } } A  \cdot \delta g^{ \mu \nu }   +  \del _{ \rho } \bigg( \sum_{ k = 0 } ^{ \infty } \D _{ \del_{ \rho } \del_{ \mu_1 } \cdots \del_{ \mu_k } f } A  \cdot  \delta \del_{ \mu_1 } \cdots \del_{ \mu_k } f \bigg) . 
\end{eqnarray}

\section{Volume term} \label{Volume term}

%%%%%%%%%%%%%%%%%%%%%%%%
Let us consider the variation of the entropy current density in terms of fundamental fields $f = \{ g^{ \rho \sig } , A_{ \alpha } , \varphi \}$ with the direction of the $d$-velocity $u^{\mu } $ fixed. To avoid unnecessary complexity, anti-symmetric second-order tensor $B_{ \mu \nu }$ is removed, and including them is straightforward. 

As seen in the last of section \ref{Therm and second law}, the first-order variation of the entropy current density is
\begin{eqnarray}
&&  \delta \Big( \sqrt{- g} u^{\mu } s \Big)   \nonumber \\
	&=&  - \lam \delta \Big( \sqrt{ - g } u^{\mu } n \Big)  +  \frac{u^{\mu } }{T}  \bigg\{  \delta \Big( \sqrt{ - g } \eps \Big) - \frac{1}{2} \sqrt{ - g } \Big[  \eps  u_{\rho } u_{\sig } + p \big( u_{ \rho } u_{ \sig }  +  g_{\rho \sig }  \big)  \Big] \delta g^{\rho \sig }  \bigg\} .  \nonumber \\  \label{varyS2}
\end{eqnarray}
If the second term of the right hand side of Eq. (\ref{varyS2}) becomes a $( d - 1 )$-dimensional total derivative of a quantity, namely 
\begin{eqnarray}
\delta \Big( \sqrt{ - g } \eps \Big) - \frac{1}{2} \sqrt{ - g } \Big[  \eps  u_{\rho } u_{\sig } + p \big( u_{ \rho } u_{ \sig }  +  g_{\rho \sig }  \big)  \Big] \delta g^{\rho \sig }  \seq  \del_a \Big( \cdots ^a \Big) \label{varyEPSstationary}
\end{eqnarray}
and the coefficients $\lam$ and $u^{ \mu } / T$ are constants in the stationary states 
\begin{eqnarray}
&&  \lam  \seq  {\rm const.}  \label{LAMconst1}  \\
&&  \frac{ u^{ \mu } }{ T }  \seq  {\rm const.} ,  \label{Tolman1}
\end{eqnarray}
then the maximum entropy principle is satisfied at least in the first-order variation. 

However, this set of the conditions (\ref{varyEPSstationary}) - (\ref{Tolman1}) is not the only one for the maximum entropy principle because the coefficients $u^{ \mu } / T$ and $\lam$ are not necessarily constants. For example, dividing $\eps$ into two parts 
\begin{eqnarray}
\eps  =  e^{ - \eta } \xi , 
\end{eqnarray}
the variation of the entropy density (\ref{varyS2}) can be arranged as 
\begin{eqnarray}
&& \delta \Big( \sqrt{- g} u^{\mu } s \Big)   \nonumber \\
	&=&  - \lam \delta \Big( \sqrt{ - g } u^{\mu } n \Big)  +  \frac{u^{\mu } }{T}  \bigg\{ \delta \Big( \sqrt{ - g } e^{ - \eta } \xi \Big)  -  \frac{ 1 }{ 2 } \sqrt{ - g } e^{ - \eta } e^{ \eta } \Big[  \eps u_{ \rho } u_{ \sig } + p \big( u_{ \rho } u_{ \sig }  +  g_{ \rho \sig } \big)  \Big] \delta g^{ \rho \sig }  \bigg\} \nonumber \\
	\ \\
	&=&    - \lam \delta \Big( \sqrt{ - g } u^{\mu } n \Big)  +  \frac{u^{\mu } }{T} e^{ - \eta }  \bigg\{ \delta \Big( \sqrt{ - g } \xi \Big)  -  \sqrt{ - g } \xi \delta \eta  -  \frac{ 1 }{ 2 } \sqrt{ - g } e^{ \eta } \Big[  \eps u_{ \rho } u_{ \sig } + p \big( u_{ \rho } u_{ \sig }  +  g_{ \rho \sig } \big)  \Big] \delta g^{ \rho \sig }  \bigg\} .  \nonumber \\  
\end{eqnarray}
By this expression, the conditions 
\begin{eqnarray}
\delta \Big( \sqrt{ - g } \xi \Big)  -  \sqrt{ - g } \xi \delta \eta  -  \frac{ 1 }{ 2 } \sqrt{ - g } e^{ \eta } \Big[  \eps u_{ \rho } u_{ \sig } + p \big( u_{ \rho } u_{ \sig }  +  g_{ \rho \sig } \big)  \Big] \delta g^{ \rho \sig }  \seq  \del_a \Big( \cdots ^a \Big)
\end{eqnarray}
and 
\begin{eqnarray}
&&  \lam  \seq  {\rm const.}  \\
&&  \frac{ u^{ \mu } }{ T } e^{ - \eta }  \seq  {\rm const.} 
\end{eqnarray}
also make the maximum entropy principle be satisfied. 

We regard the scalar field $\eta$ as a fundamental field, and every cases in which the field $\eta$ is not fundamental can be considered as a case with an extra constraint condition on $\eps$.

\subsection{From the maximum entropy principle to field equations} \label{MEPtoFE}
Let us deform $\sqrt{ - g } \eps$ as 
\begin{eqnarray}
\sqrt{ - g } \eps  =  - e^{ - \eta } u^{ \rho } u^{ \sig } \bD _{ g^{ \rho \sig } } \sqrt{ - g } \mathcal{L} , \label{decompEPS}
\end{eqnarray}
where $\mathcal{L}$ is an arbitrary function of the fundamental fields and $u^{ \mu }$. If $\eta$ is not restricted to a fundamental field, this decomposition is possible for arbitrary $\eps$, but not unique. Although in general the quantity $\eta$ is not a fundamental field, i.e., may be a function of the other fundamental fields, firstly we take $\eta$ for a independent fundamental field. The case in which $\eta$ is an general function can be obtained by imposing an extra constraint relation upon $\eta$.  

Using Eq. (\ref{varyS2}) and (\ref{decompEPS}), 
\begin{eqnarray}
&& \delta \Big( \sqrt{- g} u^{\mu } s \Big)   \nonumber \\
	&=&  - \lam \delta \Big( \sqrt{ - g } u^{\mu } n \Big)  \nonumber \\
	&&  -  \frac{u^{\mu } }{2T} e^{ - \eta } \bigg\{ 2 \delta \Big( u^{ \rho } u^{ \sig } \bD _{ g^{ \rho \sig } } \sqrt{ - g } \mathcal{L} \Big)  +  2 \sqrt{ - g } e^{ \eta } \eps \delta \eta  +  \sqrt{ - g } e^{ \eta } \Big[  \eps u_{ \rho } u_{ \sig } + p \big( u_{ \rho } u_{ \sig }  +  g_{ \rho \sig } \big)  \Big] \delta g^{ \rho \sig }  \bigg\} . \nonumber \\  \label{varyS3}
\end{eqnarray}
The Noether's identity (\ref{id3bar}) for $\mathcal{F} = \mathcal{L}$ and $f = \{ g^{ \mu \nu } , A_{ \mu } , u^{ \mu } , \eta , \varphi \}$, namely 
\begin{eqnarray}
&& 2 u^{ \rho } u^{ \nu }  \bD _{ g^{\rho \nu } } \sqrt{ - g } \mathcal{L}   -  u^{ \nu } A_{ \nu } u_{ \mu } \D _{ A_{ \mu } } \sqrt{ - g } \mathcal{L}  -  \sqrt{ - g } \mathcal{L}   \nonumber \\
&\equiv &
- u_{ \mu } u^{ \nu } \del_{\rho }  \Omega ^{\rho \mu } _{\ \ \ \nu } [ \sqrt{ - g } \mathcal{L} ]  +  \sum_{k= 0} ^{ \infty } u_{ \mu } \D_{\del_{\mu } \del_{\mu_1 } \cdots  \del_{\mu_k } f }  \sqrt{ - g } \mathcal{L} \cdot  u^{ \nu } \del_{\nu } \del_{\mu_1 } \cdots \del_{\mu_k } f . \nonumber \\ \label{id3barLag}
\end{eqnarray}
is useful for our calculation. The variation of Eq. (\ref{id3barLag}) is arranges in the form, 
\begin{eqnarray}
&&  2 \delta \Big( u^{ \rho } u^{ \sig } \bD _{ g^{ \rho \sig } } \sqrt{ - g } \mathcal{L} \Big)  \nonumber \\
	&\equiv&   \delta \Big( \sqrt{ - g } \mathcal{L} \Big)  +  A_{ \alpha } \delta \Big( u^{ \alpha } u_{ \beta } \D _{ A_{ \beta } } \sqrt{ - g } \mathcal{L} \Big)  +   u^{ \alpha } u_{ \beta } \D _{ A_{ \beta } } \sqrt{ - g } \mathcal{L} \cdot \delta A_{ \alpha }  \nonumber \\
	&&  -  \delta \Big( u_{ \mu } u^{ \nu } \del_{\rho }  \Omega ^{\rho \mu } _{\ \ \ \nu } [ \sqrt{ - g } \mathcal{L} ]  \Big)  +   \sum_{k= 0} ^{ \infty }  \bigg[  \delta \big( u_{ \mu } u^{ \nu } \big) \D_{\del_{\mu } \del_{\mu_1 } \cdots  \del_{\mu_k } f }  \sqrt{ - g } \mathcal{L} \cdot  \del_{\nu } \del_{\mu_1 } \cdots \del_{\mu_k } f  \nonumber \\
	&&  +  u_{ \mu } u^{ \nu }  \del_{ \nu } \Big( \D_{\del_{\mu } \del_{\mu_1 } \cdots  \del_{\mu_k } f }  \sqrt{ - g } \mathcal{L} \cdot  \delta \del_{\mu_1 } \cdots \del_{\mu_k } f  \Big)   \nonumber \\
	&&  +  u_{ \mu } \delta \D_{\del_{\mu } \del_{\mu_1 } \cdots  \del_{\mu_k } f }  \sqrt{ - g } \mathcal{L} \cdot  u^{ \nu } \del_{\nu } \del_{\mu_1 } \cdots \del_{\mu_k } f  -  u_{ \mu } u^{ \nu } \del_{ \nu } \D_{\del_{\mu } \del_{\mu_1 } \cdots  \del_{\mu_k } f }  \sqrt{ - g } \mathcal{L} \cdot  \delta \del_{\mu_1 } \cdots \del_{\mu_k } f   \bigg] .  \nonumber \\ \label{varyDL}
\end{eqnarray}
Substituting Eq. (\ref{varyDL}) into Eq. (\ref{varyS3}), we obtain the following expression of the variation of the entropy current: 
\begin{eqnarray}
&& \delta \Big( \sqrt{- g} u^{\mu } s \Big)   \nonumber \\
	&=&  - \lam \delta \Big( \sqrt{ - g } u^{\mu } n \Big)  -  \frac{ u^{ \mu } }{ 2 T } e^{ - \eta } A_{ \alpha } \delta \Big( u^{ \alpha } u_{ \beta } \D _{ A_{ \beta } } \sqrt{ - g } \mathcal{L}  \Big)  \nonumber \\
	&&  -  \frac{ u^{ \mu } }{ 2 T } e^{ - \eta }  \Bigg\{  \delta \Big( \sqrt{ - g } \mathcal{L} \Big)  +  2 \sqrt{ - g } e^{ \eta } \eps \delta \eta  +  \sqrt{ - g } e^{ \eta } \Big[  \eps u_{ \rho } u_{ \sig } + p \big( u_{ \rho } u_{ \sig }  +  g_{ \rho \sig } \big)  \Big] \delta g^{ \rho \sig }  \nonumber \\
	&&  +  u^{ \alpha } u_{ \beta } \D _{ A_{ \beta } } \sqrt{ - g } \mathcal{L} \cdot \delta A_{ \alpha }  -  \delta \Big( u_{ \alpha } u^{ \beta } \del_{\rho }  \Omega ^{\rho \alpha } _{\ \ \ \beta } [ \sqrt{ - g } \mathcal{L} ]  \Big)  \nonumber \\
	&&  +  u_{ \rho } u^{ \sig } \del_{ \sig } \bigg( \sum_{ k = 0 } ^{ \infty } \D _{ \del_{ \rho } \del_{ \mu_1 } \cdots \del_{ \mu_k } f } \sqrt{ - g } \mathcal{L}  \cdot  \delta \del_{ \mu_1 } \cdots \del_{ \mu_k } f  \bigg)  \nonumber \\
	&&  +  \sum_{k= 0} ^{ \infty }  \bigg[  \delta \big( u_{ \rho } u^{ \sig } \big) \D_{\del_{ \rho } \del_{\mu_1 } \cdots  \del_{\mu_k } f }  \sqrt{ - g } \mathcal{L} \cdot  \del_{ \sig } \del_{\mu_1 } \cdots \del_{\mu_k } f  \nonumber \\
	&&  +  u_{ \rho } \delta \D_{\del_{ \rho } \del_{\mu_1 } \cdots  \del_{\mu_k } f }  \sqrt{ - g } \mathcal{L} \cdot  u^{ \sig } \del_{ \sig } \del_{\mu_1 } \cdots \del_{\mu_k } f  -  u_{ \rho } u^{ \sig } \del_{ \sig } \D_{\del_{ \rho } \del_{\mu_1 } \cdots  \del_{\mu_k } f }  \sqrt{ - g } \mathcal{L} \cdot  \delta \del_{\mu_1 } \cdots \del_{\mu_k } f   \bigg]   \Bigg\} \nonumber \\ \label{varyS4}
\end{eqnarray}
Expanding $\delta ( \sqrt{ - g } \mathcal{L} )$ as 
\begin{eqnarray}
&&  \delta \Big( \sqrt{ - g } \mathcal{L} \Big)  \nonumber \\
	&=&  \bD _{ g^{ \rho \sig } } \sqrt{ - g } \mathcal{L}  \cdot  \delta g^{ \rho \sig }  +  \D_{ A_{ \alpha } } \sqrt{ - g } \mathcal{L}  \cdot  \delta A_{ \alpha }  +  \D _{ \eta } \sqrt{ - g } \mathcal{L}  \cdot  \delta \eta  +  \D_{ \varphi } \sqrt{ - g } \mathcal{L}  \cdot  \delta \varphi  \nonumber \\
	&&  +  \del_{ \rho } \bigg( \sum_{ k = 0 } ^{ \infty } \D _{ \del_{ \rho } \del_{ \mu_1 } \cdots \del_{ \mu_k } f } \sqrt{ - g } \mathcal{L}  \cdot  \delta \del_{ \mu_1 } \cdots \del_{ \mu_k } f  \bigg) , 
\end{eqnarray}
then Eq. (\ref{varyS4}) is 
\begin{eqnarray}
&& \delta \Big( \sqrt{- g} u^{\mu } s \Big)   \nonumber \\
	&=&  - \lam \delta \Big( \sqrt{ - g } u^{\mu } n \Big)  -  \frac{ u^{ \mu } }{ 2 T } e^{ - \eta } A_{ \alpha } \delta \Big( u^{ \alpha } u_{ \beta } \D _{ A_{ \beta } } \sqrt{ - g } \mathcal{L}  \Big)  \nonumber \\
	&&  -  \frac{ u^{ \mu } }{ 2 T } e^{ - \eta }  \Bigg\{  \Big(  \bD _{ g^{ \rho \sig } } \sqrt{ - g } \mathcal{L}  +  \sqrt{ - g } e^{ \eta } \big[  \eps u_{ \rho } u_{ \sig } + p ( u_{ \rho } u_{ \sig }  +  g_{ \rho \sig } )  \big]  \Big)  \delta g^{ \rho \sig }  \nonumber \\
	&&  +  \Big( \D _{ \eta } \sqrt{ - g } \mathcal{L}  +  2 \sqrt{ - g } e^{ \eta } \eps  \Big) \delta \eta  +  \big( \delta ^{ \alpha } _{ \beta }  +  u^{ \alpha } u_{ \beta } \big) \D _{ A_{ \beta } } \sqrt{ - g } \mathcal{L}  \cdot  \delta A_{ \alpha }  +  \D _{ \varphi } \sqrt{ - g } \mathcal{L}  \cdot  \delta \varphi  \nonumber \\
	&&  -  \delta \Big( u_{ \alpha } u^{ \beta } \del_{\rho }  \Omega ^{\rho \alpha } _{\ \ \ \beta } [ \sqrt{ - g } \mathcal{L} ]  \Big)  +  \big( \delta ^{ \sig } _{ \rho }  +  u^{ \sig } u_{ \rho } \big)  \del_{ \sig } \bigg( \sum_{ k = 0 } ^{ \infty } \D _{ \del_{ \rho } \del_{ \mu_1 } \cdots \del_{ \mu_k } f } \sqrt{ - g } \mathcal{L}  \cdot  \delta \del_{ \mu_1 } \cdots \del_{ \mu_k } f  \bigg)  \nonumber \\
	&&  +  \sum_{k= 0} ^{ \infty }  \bigg[  \delta \big( u_{ \rho } u^{ \sig } \big) \D_{\del_{ \rho } \del_{\mu_1 } \cdots  \del_{\mu_k } f }  \sqrt{ - g } \mathcal{L} \cdot  \del_{ \sig } \del_{\mu_1 } \cdots \del_{\mu_k } f  \nonumber \\
	&&  +  u_{ \rho } \delta \D_{\del_{ \rho } \del_{\mu_1 } \cdots  \del_{\mu_k } f }  \sqrt{ - g } \mathcal{L} \cdot  u^{ \sig } \del_{ \sig } \del_{\mu_1 } \cdots \del_{\mu_k } f  -  u_{ \rho } u^{ \sig } \del_{ \sig } \D_{\del_{ \rho } \del_{\mu_1 } \cdots  \del_{\mu_k } f }  \sqrt{ - g } \mathcal{L} \cdot  \delta \del_{\mu_1 } \cdots \del_{\mu_k } f   \bigg]   \Bigg\} . \nonumber \\ \label{varySgeneral}
\end{eqnarray}
This is the most general expression of the first-order variation of the entropy current. 

Here, we introduce the ``co-moving coordinate'' in which the $x^0$-axis is parallel to the $d$-velocity $u^{ \mu }$ so that $u^a = 0$, $u^0 u_0 = - 1$, and $\del_0 X$ for an arbitrary quantity $X$ in a stationary state. Therefore, in this coordinate, the last two lines of Eq. (\ref{varySgeneral}) vanish in a stationary state: 
\begin{eqnarray}
&& \delta \Big( \sqrt{- g} u^{\mu } s \Big)   \nonumber \\
	&=&  - \lam \delta \Big( \sqrt{ - g } u^{\mu } n \Big)  +  \frac{ u^{ \mu } }{ 2 T } e^{ - \eta } A_{ 0 } \delta \Big(  \D _{ A_{ 0 } } \sqrt{ - g } \mathcal{L}  \Big)  \nonumber \\
	&&  -  \frac{ u^{ \mu } }{ 2 T } e^{ - \eta }  \Bigg\{  \Big(  \bD _{ g^{ \rho \sig } } \sqrt{ - g } \mathcal{L}  +  \sqrt{ - g } e^{ \eta } \big[  \eps u_{ \rho } u_{ \sig } + p ( u_{ \rho } u_{ \sig }  +  g_{ \rho \sig } )  \big]  \Big)  \delta g^{ \rho \sig }  \nonumber \\
	&&  +  \Big( \D _{ \eta } \sqrt{ - g } \mathcal{L}  +  2 \sqrt{ - g } e^{ \eta } \eps  \Big) \delta \eta  +  \D _{ A_{ a } } \sqrt{ - g } \mathcal{L}  \cdot  \delta A_{ a }  +  \D _{ \varphi } \sqrt{ - g } \mathcal{L}  \cdot  \delta \varphi  \nonumber \\
	&&  +  \delta \Big(  \del_{\rho }  \Omega ^{\rho 0 } _{\ \ \ 0 } [ \sqrt{ - g } \mathcal{L} ]  \Big)  +  \del_{ a } \bigg( \sum_{ k = 0 } ^{ \infty } \D _{ \del_{ a } \del_{ \mu_1 } \cdots \del_{ \mu_k } f } \sqrt{ - g } \mathcal{L}  \cdot  \delta \del_{ \mu_1 } \cdots \del_{ \mu_k } f  \bigg)  \nonumber \\
	&&  -  \sum_{k= 0} ^{ \infty }  \bigg[  \delta \D_{ \del_{ 0 } \del_{\mu_1 } \cdots  \del_{\mu_k } f }  \sqrt{ - g } \mathcal{L} \cdot  \del_{ 0 } \del_{\mu_1 } \cdots \del_{\mu_k } f  -  \del_{ 0 } \D_{\del_{ 0 } \del_{\mu_1 } \cdots  \del_{\mu_k } f }  \sqrt{ - g } \mathcal{L} \cdot  \delta \del_{\mu_1 } \cdots \del_{\mu_k } f   \bigg]   \Bigg\}  \nonumber \\
	\\
	&\seq&    - \lam \delta \Big( \sqrt{ - g } u^{\mu } n \Big)  +  \frac{ u^{ \mu } }{ 2 T } e^{ - \eta } A_{ 0 } \delta \Big(  \D _{ A_{ 0 } } \sqrt{ - g } \mathcal{L}  \Big)  \nonumber \\
	&&  -  \frac{ u^{ \mu } }{ 2 T } e^{ - \eta }  \Bigg\{  \Big(  \bD _{ g^{ \rho \sig } } \sqrt{ - g } \mathcal{L}  +  \sqrt{ - g } e^{ \eta } \big[  \eps u_{ \rho } u_{ \sig } + p ( u_{ \rho } u_{ \sig }  +  g_{ \rho \sig } )  \big]  \Big)   \delta g^{ \rho \sig }  \nonumber \\
	&&  +  \Big( \D _{ \eta } \sqrt{ - g } \mathcal{L}  +  2 \sqrt{ - g } e^{ \eta } \eps  \Big) \delta \eta  +  \D _{ A_{ a } } \sqrt{ - g } \mathcal{L}  \cdot  \delta A_{ a }  +  \D _{ \varphi } \sqrt{ - g } \mathcal{L}  \cdot  \delta \varphi  \nonumber \\
	&&  +  \delta \Big(  \del_{\rho }  \Omega ^{\rho 0 } _{\ \ \ 0 } [ \sqrt{ - g } \mathcal{L} ]  \Big)  +  \del_{ a } \bigg( \sum_{ k = 0 } ^{ \infty } \D _{ \del_{ a } \del_{ \mu_1 } \cdots \del_{ \mu_k } f } \sqrt{ - g } \mathcal{L}  \cdot  \delta \del_{ \mu_1 } \cdots \del_{ \mu_k } f  \bigg) . \label{varyS5} \nonumber \\
\end{eqnarray}

If the field equations  
\begin{eqnarray}
	\bD _{ g^{ \rho \sig } } \sqrt{ - g } \mathcal{L}  &\seq&  - \sqrt{ - g } e^{ \eta } \big[  \eps u_{ \rho } u_{ \sig } + p ( u_{ \rho } u_{ \sig }  +  g_{ \rho \sig } )  \big]  \label{ELeqG}  \\
	\D _{ \eta } \sqrt{ - g } \mathcal{L}  &\seq&  - 2 \sqrt{ - g } e^{ \eta } \eps  \label{ELeqeta} \\
	\D _{ A_{ \rho } } \sqrt{ - g } \mathcal{L}  &\seq&  - \sqrt{ - g } u^{ \rho } q n  \label{ELeqA} \\
	\D _{ \varphi } \sqrt{ - g } \mathcal{L}  &\seq&  0 , \label{ELeqphi}
\end{eqnarray}
and the ``consistency conditions''
\begin{eqnarray}
	&& \lam  +  \frac{ u^0 }{ 2 T } e^{ - \eta } q A_0  \seq  {\rm const.} \\
	&& \frac{ u^0 }{ T } e^{ - \eta }  \seq  {\rm const.} 
\end{eqnarray}
hold for any stationary states, any variation of the fundamental fields with the conserved charges fixed and appropriate boundary conditions held does not change the total entropy, i.e., 
\begin{eqnarray}
	\delta S  &\seq&  - \bigg(  \lam  +  \frac{ u^0 }{ 2 T } e^{ - \eta } q A_0  \bigg) \delta N  +  \frac{ u^0 }{ 2 T } e^{ - \eta } \delta M_{\rm Noether}  \nonumber \\
	&&  -  \frac{ u^0 }{ 2 T } e^{ - \eta }  \oint _{ \del \mathcal{V} } ( \d ^{ d - 2 } x ) _{ 0 a }   \sum_{ k = 0 } ^{ \infty } \D _{ \del_{ a } \del_{ \mu_1 } \cdots \del_{ \mu_k } f } \sqrt{ - g } \mathcal{L}  \cdot  \delta \del_{ \mu_1 } \cdots \del_{ \mu_k } f ,  \nonumber \\
\end{eqnarray}
where $( \d ^{ d - 2 } x ) _{ \mu \nu }  :=  \frac{ 1 }{ ( d - 2 ) ! } \d x ^{\mu_1 } \wedge \cdots \wedge \d x^{\mu_{d - 2} } \epsilon_{ \mu \nu \mu_1 \cdots \mu_{d - 2} }$. From only the maximum entropy principle, one cannot know what equations hold in non-stationary states. In section \ref{consistency}, one finds that if Eqs. (\ref{ELeqG}) - (\ref{ELeqphi}) hold in not only stationary but also non-stationary states, the entropy current density is conserved, i.e., entropy production is forbidden. Therefore, such a fluid is called a perfect fluid. 

It should be noticed that the field equations (\ref{ELeqG}) - (\ref{ELeqphi}) are compatible with Eq. (\ref{decompEPS}), i.e., contracting with $- u^{ \rho } u^{ \sig }$, Eq. (\ref{ELeqG}) is
\begin{eqnarray}
	- u^{ \rho } u^{ \sig } \bD _{ g^{ \rho \sig } } \sqrt{ - g } \mathcal{L}  \seq   \sqrt{ - g } e^{ \eta } \eps , 
\end{eqnarray}
that coincides with Eq. (\ref{decompEPS}) in a stationary state. This is a nontrivial feature of $\eps$ that is described by Eq. (\ref{decompEPS}) with a fundamental scalar field $\eta$. 

The current density $ \D _{ A_{\mu } } \sqrt{ - g } \mathcal{L} = - \sqrt{ - g } u^{ \mu } q n$ should also conserve, i.e., 
\begin{eqnarray}
	\del_{\mu } \Big(  \D _{ A_{\mu } } \sqrt{ - g } \mathcal{L}  \Big)  =  0 . \label{consvDAL}
\end{eqnarray}

If one imposes the $U(1)$ gauge symmetry on the Lagrangian, the identity (\ref{id5}) shows that the conservation law (\ref{consvDAL}) requires the field $\eta$ to be electrically neutral (See appendix \ref{gauge}).

\subsection{Consistency conditions} \label{consistency}
In this subsection, we see that the field equations (\ref{ELeqG}) - (\ref{ELeqphi}) are consistent with the relations which we have assumed above: 
\begin{eqnarray}
&&  \nabla _{ \mu } \big( u ^{ \mu } s \big)  =  0 , \label{consvS2} \\
&&  \nabla _{ \mu } \big( u ^{ \mu } n \big)  =  0 , \label{consvN2} \\
&&  \frac{ u ^{ \mu } }{ T } e^{ - \eta }   \seq  {\rm const.\ in\ \mathcal{V}} , \label{etaTolman} \\
&&  \lam \delta ^{ \mu } _{ \nu }  +  \frac{ q }{ 2T } e^{- \eta }  u^{ \mu } A_{ \nu }   \seq  {\rm const.\ in\ \mathcal{V}} ,  \label{etaLAMconst}
\end{eqnarray}
where $\nabla _{ \mu }$ is the covariant derivative compatible with $g_{ \mu \nu }$. In this subsection, we use $\nabla _{ \mu }$ rather than $\del _{ \mu }$ for simpler calculation. We call Eqs. (\ref{consvS2}) - (\ref{etaLAMconst}) consistency conditions.

From Noether's theorem (\ref{id1arrange}) with replacing $\mathcal{F}$ into $\mathcal{L}$, the following identity holds: 
\begin{eqnarray}
&&  2 \sqrt{ - g } \nabla _{\nu } \bar E^{\nu } _{\mu } [ \sqrt{ - g } \mathcal{L} ]   \nonumber \\
	&\equiv&   A_{\mu } \del_{\nu } \Big( \D _{ A_{\nu } }  \sqrt{ - g } \mathcal{L} \Big)  -  F_{\mu \nu } \D _{ A_{\nu } }  \sqrt{ - g } \mathcal{L}  \nonumber \\
	&&  -  \del_{ \nu } \Big[  u^{ \nu }  \big( u_{ \mu } u^{ \rho }  +  \delta ^{ \rho } _{ \mu }  \big)  \D _{ u^{ \rho } } \sqrt{ - g } \mathcal{L}  \Big]  -  \bigg[  \del_{ \mu } u^{ \rho }  -  \frac{ 1 }{ 2 } \big( \del_{ \mu } g_{ \alpha \beta } \big) u^{ \alpha } u^{ \beta } u^{ \rho }  \bigg]  \D _{ u^{ \rho } } \sqrt{ - g } \mathcal{L}   \nonumber \\
	&&  -  ( \del_{\mu } \varphi ) \D _{ \varphi }  \sqrt{ - g } \mathcal{L}  -  ( \del_{\mu } \eta ) \D _{ \eta }  \sqrt{ - g } \mathcal{L}  , \label{id1L} \nonumber \\
\end{eqnarray}
where 
\begin{eqnarray}
E_{\mu \nu } [ \sqrt{ - g } \mathcal{L} ]  &:=&  \frac{ 1 }{ \sqrt{ - g } } \D _{ g^{\mu \nu } }  \sqrt{ - g } \mathcal{L} \\
\bar E_{\mu \nu } [ \sqrt{ - g } \mathcal{L} ]  &:=&  \frac{ 1 }{ \sqrt{ - g } } \bD _{ g^{\mu \nu } }  \sqrt{ - g } \mathcal{L} \\
	&=&  E _{ \mu \nu } [ \sqrt{ - g } \mathcal{L} ]  -  \frac{ 1 }{ 2 \sqrt{ - g } } u_{ \mu } u_{ \nu } u^{ \rho } \D _{ u^{ \rho } } \sqrt{ - g } \mathcal{L}  \\
F_{\mu \nu }  &:=&  \del_{\mu } A_{\nu }  -  \del_{\nu } A_{\mu } .  
\end{eqnarray}
Notice that the third line of Eq. (\ref{id1L}) vanishes in a stationary state. It is easier to check by direct calculation in a co-moving coordinate, in which $u^a = 0$, $u^0 =  ( - g_{00} ) ^{ - 1 / 2 }$, and $\del_0 X = 0$ for an arbitrary quantity $X$. Since the third line of Eq. (\ref{id1L}) is a covariant vector density, denoted by $- \sqrt{ - g } V_{ \mu }$, the following equation holds in any coordinates: 
\begin{eqnarray}
\sqrt{ - g } V_{ \mu }  &:=&  \del_{ \nu } \Big[  u^{ \nu }  \big( u_{ \mu } u^{ \rho }  +  \delta ^{ \rho } _{ \mu }  \big)  \D _{ u^{ \rho } } \sqrt{ - g } \mathcal{L}  \Big]  +  \bigg[  \del_{ \mu } u^{ \rho }  -  \frac{ 1 }{ 2 } \big( \del_{ \mu } g_{ \alpha \beta } \big) u^{ \alpha } u^{ \beta } u^{ \rho }  \bigg]  \D _{ u^{ \rho } } \sqrt{ - g } \mathcal{L}  \label{defV} \nonumber \\
\ \\
	&\seq&  0 . \label{vanishV}
\end{eqnarray}
In more detail, one can check as well that the parallel part to $u_{ \mu }$ vanishes in an arbitrary state: 
\begin{eqnarray}
\sqrt{ - g } u^{ \mu } V_{ \mu }  =  0 . \label{vanishUV}
\end{eqnarray}

Substituting Eqs. (\ref{ELeqG}), (\ref{ELeqphi}), and (\ref{defV}) into Eq. (\ref{id1L}), 
\begin{eqnarray}
\sqrt{ - g } \nabla _{\nu } \Big\{ e^{\eta } \Big[  \eps u^{\nu } u_{\mu }  +  p ( u^{ \nu } u_{ \mu }  +  \delta ^{\nu } _{\mu }  ) \Big]  \Big\}  =    \frac{1}{2} F_{\mu \nu } \D _{ A_{\nu } } \sqrt{ - g } \mathcal{L}  +  \frac{1}{2} ( \del_{\mu } \eta ) \D _{\eta } \sqrt{ - g } \mathcal{L}  +  \frac{ 1 }{ 2 } \sqrt{ - g } V_{ \mu }  . \label{id1L2} \nonumber \\
\end{eqnarray}
Transposing all the terms in the left hand side which contain $\eta $ to the right hand side, and separating them into parallel and orthogonal parts to $u_{ \mu } $, Eq. (\ref{id1L2}) is arranged as follows: 
\begin{eqnarray}
	&& \sqrt{ - g } \nabla _{\nu } \Big[  \eps u^{\nu } u_{\mu }  +  p \big(  u^{ \nu } u_{ \mu }  +  \delta ^{\nu } _{\mu }  \big) \Big]  \nonumber \\
	&=&  - \sqrt{ - g } ( \del_{ \nu } \eta )  \Big[  \eps u^{\nu } u_{\mu }  +  p \big(  u^{ \nu } u_{ \mu }  +  \delta ^{\nu } _{\mu }  \big) \Big]  +  \frac{1}{2} ( \del_{\mu } \eta ) e^{ - \eta } \D _{\eta } \sqrt{ - g } \mathcal{L}  +  \frac{ 1 }{ 2 } \sqrt{ - g } e^{ - \eta } V_{ \mu }  \nonumber \\
	&&  +  \frac{1}{2} e^{ - \eta } F_{\mu \nu } \D _{ A_{\nu } } \sqrt{ - g } \mathcal{L}   \\
	&=&    - u_{ \mu } u^{ \nu } \Big(  \sqrt{ - g } \eps  +  \frac{1}{2} e^{ - \eta } \D _{ \eta } \sqrt{ - g } \mathcal{L}  \Big)  \del_{\nu } \eta  \nonumber \\
	&&  -  \big( \delta ^{\nu } _{\mu }  +  u^{\nu } u_{\mu }  \big) \bigg[  \Big(  \sqrt{ - g } p  -  \frac{1}{2} e^{ - \eta } \D _{\eta } \sqrt{ - g } \mathcal{L} \Big)  \del_{\nu } \eta  -  \frac{ 1 }{ 2 } \sqrt{ - g } e^{ - \eta } V_{ \nu }   \bigg]  \nonumber \\
	&&  +  \frac{1}{2} e^{ - \eta } F_{\mu \nu } \D _{ A_{\nu } } \sqrt{ - g } \mathcal{L}  \\
	&=&  - u_{ \mu } u^{ \nu } \Big(  \sqrt{ - g } \eps  +  \frac{1}{2} e^{ - \eta } \D _{ \eta } \sqrt{ - g } \mathcal{L}  \Big)  \del_{\nu } \eta   \nonumber \\
	&&  - \big( \delta ^{\nu } _{\mu }  +  u^{\nu } u_{\mu }  \big) \bigg[  \Big(  \sqrt{ - g } p  -  \frac{1}{2} e^{ - \eta } \D _{\eta } \sqrt{ - g } \mathcal{L} \Big)  \del_{\nu } \eta  -  \frac{ 1 }{ 2 } \sqrt{ - g } e^{ - \eta } V_{ \nu }  - \frac{1}{2} e^{ - \eta }  F_{\nu \rho } \D _{ A_{\rho } } \sqrt{ - g } \mathcal{L}  \bigg] ,  \label{id1L3} \nonumber \\
\end{eqnarray}
where we use the properties Eq. (\ref{vanishUV}) and $\D _{A_{\rho } } \sqrt{ - g } \mathcal{L}  \propto u^{\rho } $. Similarly, the first line of Eq. (\ref{id1L3}) is also separated into two parts: 
\begin{eqnarray}
&& \nabla _{\nu } \Big[  \eps u^{\nu } u_{\mu }  +  p \big(  u^{ \nu } u_{ \mu }  +  \delta ^{\nu } _{\mu }  \big)   \Big]  \nonumber \\
	&=&  \Big[ u^{\nu } \del _{\nu } ( \eps + p )  +  ( \eps + p ) \nabla _{\nu } u^{\nu }  \Big] u_{\mu }  +  ( \eps + p ) u^{\nu } \nabla _{\nu } u_{\mu }  +  \del_{\mu } p   \\
	&=&  \Big[ u^{\nu } \del _{\nu } \eps  +  ( \eps + p ) \nabla _{\nu } u^{\nu }   \Big] u_{\mu }  \nonumber \\
	&&  +  ( \eps + p ) u^{\nu } \nabla _{\nu } u_{\mu }  +  \big( \delta ^{\nu } _{\mu }  +  u^{\nu } u_{\mu }  \big) \del _{\nu } p  \\
	&=&  \Big[ u^{\nu } \del _{\nu } \eps  +  ( \eps + p ) \nabla _{\nu } u^{\nu }   \Big] u_{\mu }  \nonumber \\
	&&  +  \big( \delta ^{\nu } _{\mu }  +  u^{\nu } u_{\mu }  \big)  \Big[ ( \eps + p ) u^{\rho } \nabla _{\rho } u_{\nu }  +   \del _{\nu } p \Big] . \label{decompNablaT}
\end{eqnarray}
The following calculation shows that the first term of the right hand side (\ref{decompNablaT}) is related to the conservation law of the entropy and particle number: 
\begin{eqnarray}
T \nabla _{\nu } \big( u^{\nu } s \big)  &=&  T u^{\nu } \del _{\nu } s +  T s \nabla _{\nu } u^{\nu }  \nonumber \\
	&=&  T u^{\nu } \del _{\nu } \frac{ \eps + p - \mu n }{T}  +  ( \eps + p - \mu n ) \nabla _{\nu } u^{\nu } \nonumber \\
	&=&   u^{\nu } \del _{\nu } ( \eps + p - \mu n )  -  ( \eps + p - \mu n ) \frac{ u^{\nu } \del_{\nu } T }{T}  +  ( \eps + p - \mu n ) \nabla _{\nu } u^{\nu }  \nonumber \\
	&=&   u^{\nu } \del _{\nu } \eps + ( \eps + p ) \nabla _{\nu } u^{\nu }  +  u^{\nu } \big(  \del_{\nu } p - s \del_{\nu } T -  n \del_{\nu } \mu \big)  -  \mu \nabla _{\nu } \big( u^{\nu } n \big)  \nonumber \\
	&=&  u^{\nu } \del _{\nu } \eps + ( \eps + p ) \nabla _{\nu } u^{\nu }   -  \mu \nabla _{\nu } \big( u^{\nu } n \big) , 
\end{eqnarray}
where we use the relation $\del_{\nu } p - s \del_{\nu } T -  n \del_{\nu } \mu = 0$, which is equivalent to Eq. (\ref{relationPTMu}). Thus, one obtains the relation 
\begin{eqnarray}
u^{\nu } \del _{\nu } \eps + ( \eps + p ) \nabla _{\nu } u^{\nu }  =  T \nabla _{\nu } \big( u^{\nu } s \big)  +  \mu \nabla _{\nu } \big( u^{\nu } n \big) . \label{relation1}
\end{eqnarray}
Substituting Eq. (\ref{relation1}) into (\ref{decompNablaT}), 
\begin{eqnarray}
\nabla _{\nu } \Big[ ( \eps + p ) u^{\nu } u_{\mu }  +  p \delta ^{\nu } _{\mu }   \Big]   =  \Big[ T \nabla _{\nu } \big( u^{\nu } s \big)  +  \mu \nabla _{\nu } \big( u^{\nu } n \big)  \Big] u_{\mu }  +  \big( \delta ^{\nu } _{\mu }  +  u^{\nu } u_{\mu }  \big)  \Big[ ( \eps + p ) u^{\rho } \nabla _{\rho } u_{\nu }  +   \del _{\nu } p \Big] . \label{relation2} \nonumber \\
\end{eqnarray}
Combining Eq. (\ref{id1L3}) and Eq. (\ref{relation2}), we obtain the equation 
\begin{eqnarray}
	&& \sqrt{ - g } \bigg[ T \nabla _{\nu } \big( u^{\nu } s \big)  + \mu \nabla _{\nu } \big( u^{\nu } n \big) +  \Big(  \eps  +  \frac{1}{2 \sqrt{ - g } } e^{ - \eta }  \D _{ \eta } \sqrt{ - g} \mathcal{L} \Big) u^{ \nu } \del_{\nu } \eta \bigg] u_{\mu }  \nonumber \\
	&& +   \sqrt{ - g } \big( \delta ^{\nu } _{\mu }  +  u^{\nu } u_{\mu }  \big) \bigg[ ( \eps + p ) u^{\rho } \nabla _{\rho } u_{\nu } + \del_{\nu } p  + \Big( p  -  \frac{1}{2 \sqrt{ - g } } e^{ - \eta } \D _{\eta } \sqrt{ - g } \mathcal{L} \Big) \del_{\nu } \eta  \nonumber \\
	&&  -  \frac{ 1 }{ 2 } e^{ - \eta } V_{ \nu }  -  \frac{1}{2 \sqrt{ - g } } e^{ - \eta } F_{\nu \rho } \D _{ A_{\rho } } \sqrt{ - g } \mathcal{L}  \bigg]  =  0 . \label{id1L4}
\end{eqnarray}
Since the first term of the left hand side of Eq. (\ref{id1L4}) is parallel to $u_{ \mu } $ and the second is orthogonal, both the terms should be $0$, respectively: 
\begin{eqnarray}
&& T \nabla _{\nu } \big( u^{\nu } s \big)  + \mu \nabla _{\nu } \big( u^{\nu } n \big) +  \Big(  \eps  +  \frac{1}{2 \sqrt{ - g } } e^{ - \eta }  \D _{ \eta } \sqrt{ - g} \mathcal{L} \Big) u^{ \nu } \del_{\nu } \eta  =  0 , \label{id1Lpara} \\
&& \big( \delta ^{\nu } _{\mu }  +  u^{\nu } u_{\mu }  \big) \bigg[ ( \eps + p ) u^{\rho } \nabla _{\rho } u_{\nu } + \del_{\nu } p  \nonumber \\
&& \ \ \  + \Big( p - \frac{1}{2 \sqrt{ - g } } e^{ - \eta } \D _{\eta } \sqrt{ - g } \mathcal{L} \Big) \del_{\nu } \eta  -  \frac{ 1 }{ 2 } e^{ - \eta } V_{ \nu }  - \frac{1}{2 \sqrt{ - g } } e^{ - \eta } F_{\nu \rho } \D _{ A_{\rho } } \sqrt{ - g } \mathcal{L}  \bigg]  =  0 . \label{id1Lorth} \nonumber \\
\end{eqnarray}
From Eq. (\ref{id1Lpara}), the conservation laws of $u^{ \mu } s$ and $u^{ \mu } n$ require the relation 
\begin{eqnarray}
\Big(  \eps  +  \frac{1}{2 \sqrt{ - g } } e^{ - \eta }  \D _{ \eta } \sqrt{ - g} \mathcal{L} \Big) u^{ \nu } \del_{\nu } \eta  =  0 ,  
\end{eqnarray}
that is, 
\begin{eqnarray}
\D _{ \eta } \sqrt{ - g } \mathcal{L}  =  - 2 \sqrt{ - g } e^{ \eta } \eps  
\end{eqnarray}
or 
\begin{eqnarray}
u^{ \nu } \del_{ \nu } \eta  =  0 . 
\end{eqnarray}

Next, let us consider Eq. (\ref{id1Lorth}). Using the relation among the variations $p$, $T$, and $\mu$, 
\begin{eqnarray}
\del_{ \nu } p  &=&  s \del_{ \nu } T  +  n \del_{ \nu } \mu \\
	&=&  s \del_{ \nu } T  +  n ( \lam \del_{ \nu } T + T \del_{ \nu } \lam ) \\
	&=&  ( \eps + p ) \frac{ \del_{ \nu } T }{T}  +  n T \del_{ \nu } \lam .  
\end{eqnarray}
Writing the definition of the covariant derivative explicitly, 
\begin{eqnarray}
u^{\nu } \nabla _{\nu } u_{\mu }  &=&  u^{\nu } \del _{\nu } u_{\mu }  -  u^{\nu } \Gamma ^{\rho } _{\nu \mu } u_{\rho } \\
	&=&  u^{\nu } \del _{\nu } u_{\mu }  -  \frac{1}{2} u^{\nu } u_{\rho } g^{\rho \sig } \big( \del_{\mu } g_{\sig \nu } + \del_{\nu } g_{\sig \mu }  -  \del_{\sig } g_{\nu \mu }  \big)  \\
	&=&  u^{\nu } \del_{\nu } u_{\mu }  -  \frac{1}{2} u^{\nu } u^{\sig } \del_{\mu } g_{\sig \nu } , 
\end{eqnarray}
where $\Gamma ^{ \rho } _{ \nu \mu }$ is the Christoffel symbol. In a co-moving coordinate, the $d$-velocity of a stationary fluid satisfies $u^{\rho } \del_{\rho } u_{\nu }  \seq  0$, so that 
\begin{eqnarray}
u^{ \rho } \nabla _{ \rho } u_{ \nu }  &\seq&  -  \frac{1}{2} u^{\rho } u^{\sig } \del_{\nu } g_{\rho \sig }  \\  
	&=&  -   \frac{1}{2} u^0 u^0 \del_{\nu } g_{ 00 } \\
	&=&    \frac{1}{2} \big( u^0 \big) ^2 \del_{\nu } \big( u^0 \big) ^{ - 2 }  \\
	&=&  -   \del_{ \nu } \ln u^0  , 
\end{eqnarray}
Vanishing the left hand side of Eq. (\ref{id1Lorth}) for stationary fluids in a co-moving coordinate is equivalent to 
\begin{eqnarray}
\del_a \ln u^0  -  \frac{\del_a T }{T}  -  \frac{1}{ \eps + p } \Big( p  -  \frac{1}{ 2 \sqrt{ - g } } e^{ - \eta } \D _{ \eta } \sqrt{ - g } \mathcal{L} \Big) \del_a \eta  \nonumber \\
	-  \frac{n T}{ \eps + p } \del_a \lam  +  \frac{e^{ - \eta} }{2 ( \eps + p ) } F_{ a \nu } \frac{1}{\sqrt{ - g } } \D _{ A_{\nu } } \sqrt{ - g } \mathcal{L}  \seq  0 ,  
\end{eqnarray}
where we use Eq. (\ref{vanishV}). Due to the relations $\D _{ \eta } \sqrt{ - g } \mathcal{L}  =  - 2 \sqrt{ - g} e^{ \eta } \eps $, $\D _{ A_{ \mu } } \sqrt{ - g } \mathcal{L}  =  - \sqrt{ - g } u^{ \mu } q n$, and $\del_0 A_a  =  0$, the following relation is obtained: 
\begin{eqnarray}
\bigg[  \bigg( \frac{ u^0 }{ T } e^{ - \eta }  \bigg) ^{ - 1 }  +  \frac{ q n T A_0 }{ 2 ( \eps + p ) }  \bigg]  \del_a \bigg( \frac{ u^0 }{ T } e^{ - \eta }  \bigg)  -  \frac{ n T }{ \eps + p }  \del_a  \bigg(  \lam  +  \frac{ u^0 }{ 2T } e^{- \eta }  q A_0  \bigg)   \seq  0 . \nonumber \\
\end{eqnarray}
This equation and Eq. (\ref{id1Lpara}) show that the field equations (\ref{ELeqG}) - (\ref{ELeqphi}) are compatible with the consistency conditions 
\begin{eqnarray}
&&  \nabla _{ \mu } \big( u ^{ \mu } s \big)  =  0 , \\
&&  \nabla _{ \mu } \big( u ^{ \mu } n \big)  =  0 , \\
&& \del_a  \bigg(  \frac{ u ^0 }{ T } e^{ - \eta }  \bigg)  \seq  0 , \\
&& \del_a  \bigg(  \lam  +  \frac{ u^0 }{ 2T } e^{- \eta } q A_0  \bigg)  \seq  0 . 
\end{eqnarray}

One finds the constraint condition $\eta = 0$ works well as below. Since $\delta \eta = 0$ in Eq. (\ref{varyS5}), one find the following field equations straightforwardly: 
\begin{eqnarray}
\bD _{ g^{ \rho \sig } } \sqrt{ - g } \mathcal{L}  &=&  - \sqrt{ - g } \big[  \eps u_{ \rho } u_{ \sig } + p ( u_{ \rho } u_{ \sig }  +  g_{ \rho \sig } )  \big]  \\
\D _{ A_{ \rho } } \sqrt{ - g } \mathcal{L}  &=&  - \sqrt{ - g } u^{ \rho } q n \\
\D _{ \varphi } \sqrt{ - g } \mathcal{L}  &=&  0 . 
\end{eqnarray}

\subsection{Redefinition of the physical quantities} \label{redef}
Using the field equations (\ref{ELeqeta}) and (\ref{ELeqA}), Eq. (\ref{id1Lorth}) is arranged as 
\begin{eqnarray}
\big(  u_{ \mu } u^{ \nu }  +  \delta ^{ \nu } _{ \mu }  \big) \bigg[  u^{\rho } \nabla_{\rho } u_{ \nu }  +  \frac{ \del_{ \nu } p }{ \eps + p }  +  \del_{ \nu } \eta  + \frac{ q n }{ 2 ( \eps + p ) } e^{ - \eta } F_{ \nu \rho } u^{\rho }  \bigg]   =  0 . \label{id1Lorth2}
\end{eqnarray}
Consider a situation in which $\del_a p = q = 0$ and $\del_a \eta  \neq  0$, then 
\begin{eqnarray}
u^{\rho } \nabla_{\rho } u_a  +  \del_a \eta  =  0 , 
\end{eqnarray}
in a co-moving coordinate. This equation seems strange in that the electrically neutral fluid does {\it not} flow along  geodesic orbits even when there are no gradient of the pressure, namely the fluid is force-free. 

In this subsection, we redefine the metric tensor and other physical quantities so that a force-free fluid flows along geodesic orbits.  

One finds that the new metric tensor $\t g_{ \mu \nu }$ defined by 
\begin{eqnarray}
\t g_{\mu \nu }  &:=&  e^{ 2 \eta } g_{\mu \nu } \label{defGDtilde} 
\end{eqnarray}
and $\t g^{ \mu \nu }$, $\t u^{ \mu }$, and $\t u_{ \mu }$ 
\begin{eqnarray}
\t g^{\mu \nu }  &:=&  e^{ - 2 \eta } g^{\mu \nu } \label{defGUtilde} \\
\t u^{\mu }  &:=&  e^{ - \eta } u^{\mu }  \label{defUUtilde} \\
\t u_{\mu }  &:=&  e^{ \eta } u_{\mu } \label{defUDtilde} \\
	&=&  \t g_{\mu \nu } \t u^{\nu } 
\end{eqnarray}
are appropriate for our purpose. The normalization condition of the new $d$-velocity is 
\begin{eqnarray}
\t g_{ \mu \nu } \t u^{ \mu } \t u^{ \nu } = - 1 . 
\end{eqnarray}
Indeed, $\t u^{\rho } \t \nabla _{\rho } \t u_{\nu }$ is deformed to yield 
\begin{eqnarray}
\t u^{\rho } \t \nabla _{\rho } \t u_{\nu }  &=&   u^{\rho } \t \nabla _{\rho } u_{\nu }  +  \t u^{\rho } u_{\nu } \del_{\rho } e^{ \eta } \\
	&=&  u^{\rho } \del_{\rho } u_{\nu } - \frac{1}{2} u^{\rho } \t \Gamma ^{\sig } _{ \rho \nu } u_{\sig } + u_{\nu } u^{\rho } \del_{\rho } \eta \\
	&=&  u^{\rho } \del_{\rho } u_{\nu } - \frac{1}{2} u^{\rho } u_{\sig } \t g^{\sig \omega } \big( \del_{\nu } \t g_{\omega \rho } + \del_{\rho } \t g_{\omega \nu } - \del_{\omega } \t g_{\rho \nu }  \big) + u_{\nu } u^{\rho } \del_{\rho } \eta  \\
	&=&  u^{\rho } \del_{\rho } u_{\nu } - \frac{1}{2} u^{\rho } u_{\sig }  g^{\sig \omega } \del_{\nu } g_{ \omega \rho }  -  u^{\rho } u_{ \sig } \t g^{\sig \omega } g_{\omega \rho } e^{ 2 \eta } \del_{\nu } \eta  + u_{\nu } u^{\rho } \del_{\rho } \eta \nonumber \\
	&=&  u^{\rho } \nabla _{\rho } u_{\nu }  +  \del_{\nu } \eta  + u_{\nu } u^{\rho } \del_{\rho } \eta \\
	&=&  u^{\rho } \nabla _{\rho } u_{\nu }  +  \big( \delta ^{ \rho } _{ \nu }  +  u_{ \nu } u^{ \rho } \big)  \del_{ \rho } \eta , 
\end{eqnarray}
so that Eq. (\ref{id1Lorth2}) is 
\begin{eqnarray}
\big( \delta^{\nu } _{\mu } + \t u^{\nu } \t u_{\mu } \big)  \bigg[ \t u^{\rho } \t \nabla_{\rho } \t u_{\nu }  +  \frac{ \del_{\nu } p }{ \eps + p }   + \frac{ q n }{ 2 ( \eps + p ) }  F_{\nu \rho } \t u^{\rho }   \bigg]  =  0 . \label{id1Lorthtilde}
\end{eqnarray}
In the situation $\del_{ \nu } p = q = 0$, 
\begin{eqnarray}
\t u^{\rho } \t \nabla_{\rho } \t u_{\nu }  =  0 . \label{geodesic}
\end{eqnarray}
Therefore, with constant pressure, the neutral fluid flows geodesic orbits defined by the new metric tensor $\t g_{ \mu \nu } $. Notice that the relation (\ref{geodesic}) also holds for the metric tensor $\t g_{ \mu \nu }  =  e^{ 2 ( \eta + a ) } g_{\mu \nu }$, where $a$ is an arbitrary constant. 

The determinant of $\t g_{ \mu \nu } $ is 
\begin{eqnarray}
\sqrt{ - g }  =  e^{ - d \eta } \sqrt{ - \t g } . 
\end{eqnarray}

The extensive quantities $s$, $n$ and $\eps $ are redefined by 
\begin{eqnarray}
\t v &:=&  e^{ - (d - 1) \eta }  \label{defVtilde} \\
\t s  &:=&  s \t v  \label{defStilde} \\
\t n  &:=&  n \t v  \label{defNtilde} \\
\t \eps  &:=&  \eps \t v , \label{defEPStilde}
\end{eqnarray}
so that their current densities are invariant under the redefinition: 
\begin{eqnarray}
\sqrt{ - \t g } \t u^{ \mu }  \t s  &=&  \sqrt{ - g } u^{ \mu }  s \\
\sqrt{ - \t g } \t u^{ \mu }  \t n  &=&  \sqrt{ - g } u^{ \mu }  n .   
\end{eqnarray}
The new Lagrangian $\mathcal{ \t L}$ is defined in order to make the Lagrangian density $\sqrt{ - g } \mathcal{L}$ be invariant: 
\begin{eqnarray}
\mathcal{\t L}  &:=&  e^{ - d \eta }  \mathcal{L} . 
\end{eqnarray}

The relation between $\eps$ and fundamental fields, Eq. (\ref{decompEPS}), transforms into 
\begin{eqnarray}
\sqrt{ - \t g } \t \eps  =  - \t u^{ \mu } \t u^{ \nu } \D _{ \t g^{ \mu \nu } } \sqrt{ - \t g } \mathcal{\t L} . 
\end{eqnarray}

From Eq. (\ref{defStilde}) - (\ref{defEPStilde}), the first law of thermodynamics and Gibbs-Duhem relation are modified: 
\begin{eqnarray}
\delta \t \eps  =  T \delta \t s  -  p \delta \t v  +  \mu \delta \t n \\
T \t s  +  \mu \t n  =  \t \eps  +  p \t v . 
\end{eqnarray}

The equation (\ref{id1Lorthtilde}) is rewritten as 
\begin{eqnarray}
\big( \delta^{\nu } _{\mu } + \t u^{\nu } \t u_{\mu } \big)  \bigg[ \t u^{\rho } \t \nabla_{\rho } \t u_{\nu }  +  \frac{ \t v \del_{\nu } p }{ \t \eps + p \t v }   + \frac{ q \t n }{ 2 ( \t \eps + p \t v ) }  F_{\mu \rho } \t u^{\rho }   \bigg]  =  0 . 
\end{eqnarray}

We have two kinds of geometry defined by two metric tensors $g_{ \mu \nu } $ and $\t g_{ \mu \nu } $, respectively, those are related by the conformal transformation (\ref{defGDtilde}). While the former, that we call the thermodynamical geometry, makes the description of the thermodynamics be simple, the latter, we call the kinematical geometry, does that of kinematics of the fluid be simple. The quantity $\t v$, equivalently $\eta $, represents the conformal factor between the two spatial volume elements $\sqrt{ h }$ and $\sqrt{ \t h }$ defined by $g_{\mu \nu } $ and $\t g_{\mu \nu } $, respectively, i.e.,  
\begin{eqnarray}
\sqrt{ h }  =  \t v \sqrt{ \t h } . 
\end{eqnarray}
We can interpret $\t v$ as the thermodynamical volume element measured by kinematical one.

Let us rewrite the variation of the entropy current density. The variation of $g^{\mu \nu } $ is separated as 
\begin{eqnarray}
\delta g^{\mu \nu }  &=&  \delta \big( e^{ 2 \eta } \t g^{\mu \nu }  \big)  \\
	&=&  e^{ 2 \eta } \delta \t g^{\mu \nu }  +  2 e^{ 2 \eta } \t g^{\mu \nu } \delta \eta . \label{varyG}
\end{eqnarray}
Therefore, substituting Eq. (\ref{varyG}) into Eq. (\ref{varyS5}), 
\begin{eqnarray}
&&  \delta \Big( \sqrt{ - \t g } \t u^0 \t s \Big)  +  \t \lam_0  \delta \Big( \sqrt{ - \t g } \t u^0 \t n \Big) \nonumber \\
	&\seq&   - \frac{u^0 }{2 T} e^{ - \eta } 
\bigg\{
	\delta \Big(  \sqrt{ - \t g } \mathcal{\t L}  \Big)  -  \D _{ A_0 } \sqrt{ - \t g } \mathcal{\t L}  \cdot  \delta A_0  +  2 ( d - 1 ) \sqrt{ - \t g } p \t v \delta \eta  \nonumber \\
	&&  +  \sqrt{ - \t g }  \Big[  \t \eps  \t u_{ \rho } \t u_{ \sig } +  p \t v \big(  \t u_{ \rho } \t u_{ \sig } + \t g_{\rho \sig }  \big)  \Big]  \delta \t g^{\rho \sig }  +   \delta \Big(  \del_{ \rho } \t \Omega ^{ \rho 0 } _{ \ \ \ 0 } [ \sqrt{ - \t g } \mathcal{\t L} ] \Big) \nonumber \\
	&&  +  \del_a \bigg( \sum_{ k = 0 } ^{ \infty }  \D _{ \del_a \del_{ \mu_1 } \cdots \del_{ \mu_k } \t f } \sqrt{ - \t g } \mathcal{\t L}  \cdot  \delta \del_{ \mu_1 } \cdots \del_{ \mu_k } \t f  \bigg)  
\bigg\} ,  \label{varySbar}
\end{eqnarray}
where
\begin{eqnarray}
\t \lam _0  =  \frac{ \mu }{T}  +  \frac{ q }{ 2 T } \t u ^0 A_0 ,  
\end{eqnarray}
$\t f  =  \{ \t g^{ \rho \sig } , A_{ \mu } , \eta , \varphi  \}$, and $\t \Omega ^{ \rho 0 } _{\ \ \ 0}$ is defined by Eq. (\ref{defOmega}) with $g^{ \rho \sig }$ and $u^{ \mu }$ replaced by $\t g^{ \rho \sig }$ and $\t u^{ \mu }$, respectively. The following field equations are derived by the maximum entropy principle:  
\begin{eqnarray}
\bD _{ \t g ^{\mu \nu } }  \sqrt{ - \t g } \mathcal{ \t L }  &=&   - \sqrt{ - \t g } \Big[ \t \eps  \t u_{\mu } \t u_{\nu } + p \t v \big(  \t u_{ \mu } \t u_{ \nu }  +  \t g_{\mu \nu }  \big)  \Big]  \label{ELeqGtilde} \\
\D _{ A_{\mu } } \sqrt{ - \t g } \mathcal{ \t L }  &=&  - \sqrt{ - \t g } q  \t n \t u ^{\mu } \label{ELeqAtilde} \\
\D _{ \eta }  \sqrt{ - \t g } \mathcal{ \t L }  &=&   - 2 ( d - 1 ) \sqrt{ - \t g } p \t v \label{ELeqetatilde} \\
\D _{\varphi } \sqrt{ - \t g } \mathcal{ \t L}  &=&  0 . \label{ELeqphitilde}
\end{eqnarray}
The consistency conditions are 
\begin{eqnarray}
\frac{ \t u^0 }{T}   \seq  {\rm const.} \label{Tolmanbar} \\
\t \lam_0  \seq  {\rm const.} . 
\end{eqnarray}
The condition (\ref{Tolmanbar}) coincides with Tolman's law in the kinematical geometry.

\section{Surface term} \label{Surface term}
In section \ref{Volume term}, we see that the field equations (\ref{ELeqGtilde}) - (\ref{ELeqphitilde}) maximize the total entropy $S$ in appropriate boundary conditions. When a configuration satisfies the field equations, the variation of the entropy is written in the following form: 
\begin{eqnarray}
\delta S  &\seq&   - \bigg(  \lam  +  \frac{ \t u^0 }{ 2 T } A_0  \bigg) \delta N  +  \frac{ \t u^0 }{ 2 T }  \delta \t M_{\rm Noether}  \nonumber \\
	&&  -  \frac{ \t u^0 }{ 2 T }  \oint _{ \del \mathcal{V} } ( \d ^{ d - 2 } x ) _{ 0 a }   \sum_{ k = 0 } ^{ \infty }  \D _{ \del_{ a } \del_{ \mu_1 } \cdots \del_{ \mu_k } \t f } \sqrt{ - \t g } \mathcal{ \t L}  \cdot  \delta \del_{ \mu_1 } \cdots \del_{ \mu_k } \t f , \nonumber \\
\end{eqnarray}
where the quantities 
\begin{eqnarray}
\D _{ \del_{ a } \del_{ \mu_1 } \cdots \del_{ \mu_k } \t f } \sqrt{ - \t g } \mathcal{\t L}
\end{eqnarray}
can be regarded as the chemical potentials\footnote{Notice that the variations $\delta \t f \big| _{ \del \mathcal{V} }$ and $\delta \del_{ i_1 } \cdots \del_{ i_k } \t f \big| _{ \del \mathcal{V} }$ are not independent. } conjugate to $\del_{ \mu_1 } \cdots \del_{ \mu_k } \t f$.

In this section, as an example, we calculate the variation of the entropy in the Lovelock theory including the surface terms, and see that the variation can be represented by those of the total Kodama energy and area of the boundary. For simplicity, in this section, the tilde ``$\tilde \ $'' is omitted.

\subsection{Spherically symmetric spacetime and the generalized Kodama flux}
In the Lovelock theory, it is convenient to describe the spherically symmetric system by the warped product coordinate (\ref{warped product}) due to the relation between the Lovelock tensors and Kodama flux. In this subsection, we introduce the warped product coordinate and briefly summarize the properties of the (generalized) Kodama flux \cite{Kodama1980, AV2010}.

In any spherically symmetric spacetime, there are coordinates such that the line element $\d l$ is written in the form 

\begin{eqnarray}
\d l ^2  &=&  g_{ \mu \nu } \d x^{ \mu } \d x^{ \nu } \\
	&=&  ^{\mathrm B} \!\! g_{ \eta \xi } \d x^{ \eta } \d x^{ \xi } + \chi ( r )^2 \  ^{\mathrm F} \!\! g_{ i j } \d x^i \d x^j ,  \label{warped product}
\end{eqnarray}
where $\eta$ and $\xi$ run over $\{ 0, \, 1 \}$, while $i$ and $j$ do $\{ 2, \, \dots, \, d - 1 \}$, and $r$ is the radial component of the coordinate. In the restricted coordinate, one can see the total spacetime $( \mathcal{M}^d, \, g_{ \mu \nu } )$ as a ``warped product'' manifold with the $2$-dimensional ``base space'' $( \, ^{\mathrm B} \! \mathcal{M} ^2 , \, \Bg _{ \eta \xi } )$ and the $( d - 2 )$-dimensional ``fibers'' $( ^{\mathrm F} \! \mathcal{M} ^{ d - 2 } , \, \Fg _{ i j } )$. The scalar function on the base space $\chi ( r )$ is called the ``warp factor''. 

The square root of the determinant of the metric $\sqrt{ - g }$ is decomposed as 
\[
\sqrt{ - g }  =  \chi ^{d - 2} \sqrt{ \Fg } \sqrt{ - ^{\mathrm B}\! g } . 
\]

Define the following quantity: 
\begin{eqnarray}
\epsilon ^{ \mu \nu }  &:=&  
	\left(  
		\begin{array}{cc}
		 ^{\mathrm B} \! \epsilon ^{ \eta \xi } & 0 \\
		 0 & 0 
		\end{array}
	\right)  \\
&=&  
	\left(  
		\begin{array}{cccc}
		0 & 1 & 0 & \ldots \\
		-1 & 0 & 0 & \ldots \\
		0 & 0 & 0 & \ldots \\
		\vdots & \vdots & \vdots & \ddots 
		\end{array}
	\right) , 
\end{eqnarray}
where $^{\mathrm B} \! \epsilon ^{ \eta \xi }$ denotes the Levi-civita symbol on the base space, that transforms as a tensor density of weight $1$ on the base space, and $\epsilon ^{ \mu \nu }$ transforms as a tensor density of weight $1$ on the total spacetime under the restricted coordinate transformation. The quantity 
\begin{eqnarray}
^{\mathrm B} \! e^{ \eta \xi }  :=  \frac{ ^{\mathrm B} \! \epsilon ^{ \eta \xi } }{ \sqrt{ - ^{\mathrm B} \! g } }
\end{eqnarray}
is the Levi-Civita tensor on the base space.  

Note that under the restricted coordinate transformation, there are more fields that transform as tensor fields than ordinary tensor fields. In this section, the action of the covariant derivative $\nabla _{ \mu }$ on the generalized tensor fields are defied by the same way as the ordinary ones.  

Consider the following covariant derivative: 
\begin{eqnarray}
\nabla _{ \mu }  \bigg(   \frac{1 }{ \chi ^{ d - 2 } }  \frac{ \epsilon ^{ \mu \nu } }{ \sqrt{ - \Bg } }  \bigg)  &=&  \frac{ 1 }{ \sqrt{ - g } }  \del_{ \mu }  \bigg( \frac{ \sqrt{ - g } }{ \chi ^{ d - 2 } }  \frac{ \epsilon ^{ \mu \nu } }{ \sqrt{ - \Bg } } \bigg) \\
	&=&  \frac{ 1 }{ \sqrt{ - g } }  \del_{ \mu } \Big( \sqrt{ \Fg }  \epsilon ^{ \mu \nu } \Big) \\
	&=&  \frac{ \sqrt{ \Fg } }{ \sqrt{ - g } }  \del_{ \mu } \epsilon ^{ \mu \nu }  \\
	&=&  0 . 
\end{eqnarray}
Therefore, defining $e^{ \mu \nu }  :=  \epsilon ^{ \mu \nu } / \sqrt{ - \Bg } $, the quantity 
\begin{eqnarray}
\frac{ e^{ \mu \nu } }{ \chi ^{ d - 2 } } 
\end{eqnarray}
is a covariantly divergence-free second-order anti-symmetric (generalized) tensor field. Using these features and an arbitrary (generalized) scalar field $\Psi$, one can construct a current $J _{ [ \Psi ] } ^{ \mu } $, called the generalized Kodama flux: 
\begin{eqnarray}
J_{ [ \Psi ] } ^{ \mu }  &:=&   \frac{e^{ \mu \nu } }{ \chi ^{d - 2} } \nabla _{ \nu } \Psi \\
	&=&  \sqrt{ \Fg }  \frac{ \epsilon^{ \mu \nu} }{ \sqrt{ - g } } \del_{ \nu } \Psi . \label{defPsi}
\end{eqnarray}
Due to the covariantly divergent-freeness of $e^{ \mu \nu } / \chi ^{ d - 2 }$, the divergence of $J_{ [ \Psi ] } ^{ \mu } $ vanishes, i.e., 
\begin{eqnarray}
\nabla _{\mu } J_{ [ \Psi ] } ^{ \mu }  &=&  \frac{e^{ \mu \nu } }{ \chi ^{d - 2} } \nabla _{ \mu } \nabla _{ \nu } \Psi \\
	&=&  0 . 
\end{eqnarray}

For $\Psi = \frac{ 1 }{ d - 1 } \chi ^{ d - 1 }$, 
\begin{eqnarray}
k^{ \mu }  &:=&  J^{ \mu } _{ [ \frac{ 1 }{ d - 1 } \chi ^{ d - 1 } ] }  \\ 
	&=&  \sqrt{ \Fg } \frac{ \epsilon ^{ \mu \nu } }{ \sqrt{ - g } } \chi ^{ d - 2 } \del_{ \nu } \chi  \\
	&=&  \frac{ \epsilon ^{ \mu \nu } }{ \sqrt{ - \Bg } } \del_{ \nu } \chi  \\
	&=&  e^{ \mu \nu } \del_{ \mu } \chi 
\end{eqnarray}
is called the ``Kodama vector''.

\subsection{The Lovelock theory}
In this subsection, we consider the Lovelock theory of gravity in the warped product coordinate (\ref{warped product}). The Lagrangian of the Lovelock theory \cite{Lovelock1971} is  
\begin{eqnarray}
\mathcal{L}_{\rm Love.}   &=&   \frac{ 1 }{ \kappa ^2 }  L_{\rm Love.} \\
L_{\rm Love.}   &:=&   \sum _{ n = 0 } ^{ [ d / 2 ] } a _{ ( n ) } L_{ ( n ) } \\
L _{ ( n ) }  &:=&  \frac{1}{ 2^n }  \delta ^{ \mu_1 \cdots \mu_n \nu_1 \cdots \nu_n } _{ \rho_1 \cdots \rho_n \sig_1 \cdots \sig_n } R_{ \mu_1 \nu_1 } ^{ \ \ \ \ \ \ \rho_1 \sig_1 } \cdots R_{ \mu_n \nu_n } ^{ \ \ \ \ \ \ \rho_n \sig_n } , 
\end{eqnarray}
where 
\begin{eqnarray}
&& \delta ^{ \mu_1 \cdots \mu_n } _{ \rho_1 \cdots \rho_n }  :=  n! \delta ^{ \mu_1 } _{ [ \nu_1 } \cdots \delta ^{ \mu_n } _{ \nu_n ] } , \\
&& R^{ \mu } _{ \ \ \nu \rho \sig }  :=  \del_{ \rho } \Gamma ^{ \mu } _{ \nu \sig }  -  \del_{ \sig } \Gamma ^{ \mu } _{ \nu \rho }  +  \Gamma ^{ \mu } _{ \alpha \rho } \Gamma ^{ \alpha } _{ \nu \sig }  -  \Gamma ^{ \mu } _{ \alpha \sig } \Gamma ^{ \alpha } _{ \nu \rho }  , 
\end{eqnarray}
and $a_{ ( n ) }$, $ n = 0 , \, 1 , \, \cdots , \, [ d / 2 ]$ are the parameters of the theory. The parameter $a_{ ( 0 ) }$ is related to the cosmological constant $\Lambda$ by $a_{ ( 0 ) } = - 2 \Lambda$. The Euler-Lagrange derivative of $\sqrt{ - g } \mathcal{L}_{\rm Love.}$ is 
\begin{eqnarray}
G^{ \mu } _{\ \  \nu }   &:=&  g^{ \mu \rho } \D _{ g^{ \rho \nu } } \sqrt{ - g } \mathcal{L}_{\rm Love.}  \\
	&=&  \sum_{ n = 0 } ^{ [ n / 2 ] } a_{ ( n ) } G_{ ( n ) \  \nu } ^{ \ \ \  \mu } ,  
\end{eqnarray}
where
\begin{eqnarray} 
G_{ ( n ) \ \nu } ^{ \ \ \ \mu }  &=&  - \frac{ 1 }{ 2 ^{ n + 1 } } \delta ^{\mu \mu_1 \cdots \mu_n \rho_1 \cdots \rho_n } _{ \nu \nu_1 \cdots \nu_n \sig_1 \cdots \sig_n } R_{ \mu_1 \rho_1 } ^{ \ \ \ \ \ \ \nu_1 \sig_1 } \cdots R_{ \mu_n \rho_n } ^{ \ \ \ \ \ \ \nu_n \sig_n }  . 
\end{eqnarray}
The tensor $G_{ ( n ) \ \nu } ^{ \ \ \ \mu }$ is called the $n$-th order Lovelock tensor. %The field equation of the Lovelock theory with a perfect fluid is 
%\begin{eqnarray}
%\frac{ 1 }{ \kappa ^2 } G^{ \mu } _{ \ \  \nu }  =  \eps u^{ \mu } u_{ \nu }  +  p \big( u^{ \mu } u_{ \nu }  +  \delta ^{ \mu } _{ \nu } \big) . 
%\end{eqnarray}
%In co-moving coordinates, 
%\begin{eqnarray}
%\sqrt{ - g } \eps   &=&   - \frac{ 1 }{ \kappa ^2 } \sqrt{ - g } G^0 _{ \ \ 0 }  . 
%\end{eqnarray}

In a spherically symmetric spacetime with a warped product structure of the metric (\ref{warped product}), the $( \eta, \, \xi )$ components of the $n$-th order Lovelock tensor is written in the form \cite{MWR2011} 
\begin{eqnarray}
G_{ ( n ) \ \xi } ^{ \ \ \ \eta }  &=&  - \frac{ n ( d - 2 ) ! }{ ( d - 1 - 2 n ) ! } \frac{ ^{\mathrm B} \! \nabla^{ \eta } \! \, ^{\mathrm B} \! \nabla_{ \xi } \chi  - ( ^{\mathrm B} \! \nabla ^{ \zeta } \! \, ^{\mathrm B} \! \nabla _{ \zeta } \chi ) \delta ^{ \eta } _{ \xi }  }{ \chi } \bigg( \frac{ k_{\mathrm F} - | ^{\mathrm B} \! \nabla \chi |^2 }{ \chi ^2 } \bigg) ^{ n - 1 }  \nonumber \\
	&&  -  \frac{ ( d - 2 ) ! }{ 2 ( d - 2 - 2 n ) ! } \delta ^{ \eta } _{ \xi } \bigg( \frac{ k_{\mathrm F} - | ^{\mathrm B} \! \nabla \chi |^2 }{ \chi ^2 } \bigg) ^n , 
\end{eqnarray}
where $\eta , \, \xi = 0 , \, 1$ and $| \Bnabla \chi | ^2 := \Bg ^{ \eta \xi } \Bnabla _{ \eta } \chi \Bnabla _{ \xi } \chi$. Consider the quantity 
\begin{eqnarray}
G_{ ( n ) \ \xi } ^{ \ \ \ \eta } k^{ \xi }  &=&  \sqrt{ \Fg }  \frac{ \epsilon ^{ \eta \xi } }{ \sqrt{ - g } }  \frac{ ( d - 2 ) ! }{ 2 ( d - 1 - 2 n ) ! } \Big[ n \chi ^{ d - 1 - 2 n } \del_{ \xi } \big( | \Bnabla \chi | ^2 \big)  \big( k_{\mathrm F}  -  | \Bnabla \chi | ^2 \big) ^{ n - 1 } \nonumber \\
	&&  -  ( d - 1 - 2 n ) \chi ^{ d - 2 - 2 n } \del_{ \xi } \chi  \big( k_{\mathrm F}  -  | \Bnabla \chi | ^2 \big) ^n   \Big]  \\
	&=&   \sqrt{ \Fg } \frac{ \epsilon ^{ \eta \xi } }{ \sqrt{ - g } } \del_{ \xi } \bigg[ -   \frac{ ( d - 2 ) ! }{ ( d - 1 - 2 n ) ! } \frac{ \chi ^{ d - 1 - 2 n } }{ 2 } \big( k_{\mathrm F}  -  | \Bnabla \chi | ^2 \big) ^n \bigg]  \\
	&=&   \sqrt{ \Fg } \frac{ \epsilon ^{ \eta \xi } }{ \sqrt{ - g } } \del_{ \xi } \bigg(  - \frac{ \kappa ^2 }{ A_{ \mathrm F } } m_{ ( n ) }  \bigg) , 
\end{eqnarray}
where 
\begin{eqnarray}
m_{ ( n ) }  :=  \frac{ A _{\mathrm F} ( d - 2 ) ! }{ 2 \kappa ^2 ( d - 1 - 2 n ) ! } \chi ^{ d - 1 - 2 n } \big( k_{\mathrm F}  -  | \Bnabla \chi | ^2 \big) ^n . 
\end{eqnarray}
Define the following quantity $m$, called the generalized Misner-Sharp mass \cite{MWR2011}, by 
\begin{eqnarray}
m  &:=&  \sum _{ n = 0 } ^{ [ d / 2 ] } a _{ ( n ) } m_{ ( n ) } \\
	&=&  \sum_{ n = 0 } ^{ [ d / 2 ] }  \frac{ a_{ ( n ) } A_{ \mathrm F } ( d - 2 ) ! }{ 2 \kappa ^2 ( d - 1 - 2 n ) ! } \chi ^{ d - 1 - 2 n } \big( k_{\mathrm F}  -  | \Bnabla \chi |^2  \big) ^n  \\
	&\sseq&  \sum _{ n = 0 } ^{ [ d / 2 ] } b_{ ( n ) } \chi ^{ d - 1 - 2 n }  \bigg( k_{\mathrm F}  -  \frac{ ( \del_1 \chi ) ^2 }{ g_{11} } \bigg) ^n ,  \label{defm}
\end{eqnarray}
and
\begin{eqnarray}
b_{ ( n ) }  :=  \frac{ a_{ ( n ) } A_{ \mathrm F } ( d - 2 ) ! }{ 2 \kappa ^2 ( d - 1 - 2 n ) ! } . 
\end{eqnarray}
The equation (\ref{defm}) relates $m$, $\chi$, $\del_1 \chi$, and $m$. Note that since in Eq. (\ref{defm}), $g_{11}$ and $\del_1 \chi$ appear only in the form $( \del_1 \chi ) ^2 / g_{11}$, the quantity $( \del_1 \chi ) ^2 / g_{11}$ depends only on $m$ and $\chi$: 
\begin{eqnarray}
\frac{ ( \del_1 \chi ) ^2 }{ g_{11} }  =  W( m , \chi ) ,  \label{defW}
\end{eqnarray}
where $W( m , \chi )$ is a function of $m$ and $\chi$. From Eq. (\ref{defW}), the partial derivative of $g_{11}$ with respect to $\del_1 \chi$ is obtained: 
\begin{eqnarray}
\frac{ \del g_{11} }{ \del \del_1 \chi }  =  \frac{ 2 g_{11} }{ \del_1 \chi } \\
\frac{ \del g^{11} }{ \del \del_1 \chi }  =  - \frac{ 2 g^{11} }{ \del_1 \chi } . 
\end{eqnarray}

When the coefficients $a_{ ( n ) }$ are constants, the contraction of the Lovelock tensor and the Kodama vector is arranged in the form 
\begin{eqnarray}
G ^{ \eta } _{ \ \ \xi } k^{ \xi }  &=&  \sqrt{ \Fg } \frac{ \epsilon ^{ \eta \xi } }{ \sqrt{ - g } } \del_{ \xi } \bigg( - \frac{ \kappa ^2 }{ A _{ \mathrm F } } m \bigg)  . 
\end{eqnarray}
Comparing with the definition of the generalized Kodama flux (\ref{defPsi}), 
\begin{eqnarray}
G ^{ \ \ \  \mu } _{ ( n ) \ \nu } k^{ \nu }   &=&   J_{ [ - \kappa ^2 m_{ ( n ) } / A_{ \mathrm F } ] } ^{ \mu }  \\
	&=:&  J_{ ( n ) } ^{ \mu } \\
G ^{ \mu } _{ \ \  \nu } k^{ \nu }   &=&   J_{ [ - \kappa ^2 m / A_{ \mathrm F } ] } ^{ \mu } \\
	&=:&  J^{ \mu } . 
\end{eqnarray}
The conserved current $J^{ \mu }$ is called the Kodama current for the Lovelock theory \cite{Kodama1980}.

\subsection{The variation of the entropy in the Lovelock theory}
Here, we calculate the variation of the entropy of a spherically symmetric and static neutral fluid system using the special form of the Lovelock tensor.

In a co-moving coordinate, the energy density $\sqrt{ - g } \eps$ is arranged as 
\begin{eqnarray}
\sqrt{ - g } \eps  &=&   \frac{ 1 }{ \kappa^2 } \sqrt{ - g } G^0 _{ \ \ 0 } \\
	&\sseq&   \frac{ \sqrt{ - g } }{ A_{ \mathrm F } \chi ^{ d - 2 } \del_1 \chi } \del_1 m \\
	&=&  \frac{ \sqrt{ \Fg } \sqrt{ - \Bg } \del_1 m }{ A_{ \mathrm F } \del_1 \chi }  \\
	&=&  \frac{ \sqrt{ \Fg } \sqrt{ - g_{ 00 } } \sqrt{ g_{ 11 } }  \del_1 m }{ A_{ \mathrm F } \del_1 \chi  } \\
	&=&  \frac{ \sqrt{ \Fg } \sqrt{ - g_{ 00 } }  \del_1 m }{ A_{ \mathrm F } \sqrt{ W( m , \chi ) }  } . \label{LoveEPS}
\end{eqnarray}

Let us employ $f = \{ g_{00} , \, m , \, \chi \}$ as the fundamental fields. We see in Eq. (\ref{LoveEPS}) that $\sqrt{ - g } \eps$ does not depend on neither $\del_1 g_{00}$, $\del_1 \del_1 m$, $\del_1 \chi$ nor their higher derivatives. Therefore, the variation of the entropy is 
\begin{eqnarray}
\delta S  &\sseq&  \int _{ \mathcal{V} } \d ^{ d - 1 } x  \frac{ u^0 }{ T } 
	\bigg\{  
	\D _{ g_{00} } \sqrt{ - g } \eps  \cdot \delta g_{00}  +  \D _m \sqrt{ - g } \eps \cdot \delta m  +  \D _{ \chi } \sqrt{ - g } \eps  \cdot \delta \chi \nonumber \\
	&& \ \ \ \ \ \  -  \frac{ 1 }{ 2 }  \sqrt{ - g } \Big[ \eps u^{ \mu } u^{ \nu }  +  p \big( u^{ \mu } u^{ \nu }  +  g^{ \mu \nu } \big) \Big] \delta g_{ \mu \nu } 
	\bigg\}  \nonumber \\
&& +  \oint _{ \del \mathcal{V} } \d ^{ d - 2 } x  \frac{ u^0 }{ T }  \bigg(  \D _{ \del_1 m } \sqrt{ - g } \eps \cdot \delta m  \bigg) . \label{LovevaryS}
\end{eqnarray}
From the expression (\ref{LoveEPS}), the Euler-Lagrange derivatives of $\sqrt{ - g } G^0 _{ \ \ 0 }$ are easily obtained: 
\begin{eqnarray}
\D _{ \del_1 m } \sqrt{ - g } \eps  &=&  \frac{ 1 }{ \kappa ^2 } \D _{ \del_1 m } \sqrt{ - g } G ^0 _{ \ \  0 }  \\
	&=&  \frac{ \sqrt{ \Fg } \sqrt{ - \Bg } }{ A_{ \mathrm F } \del_1 \chi }  \\
	&=&   \frac{ \sqrt{ \Fg } \sqrt{ - g_{ 00 } } \sqrt{ g_{ 11 } } }{ A_{ \mathrm F } \del_1 \chi } . 
\end{eqnarray}

The second line of Eq. (\ref{LovevaryS}) is arranged as 
\begin{eqnarray}
&&    \frac{ 1 }{ 2 }  \sqrt{ - g } \Big[ \eps u^{ \rho } u^{ \sig }  +  p ( u^{ \rho } u^{ \sig }  +  g^{ \rho \sig } )  \Big]  \delta g_{ \rho \sig }  \\
	&\sseq&    \frac{ 1 }{ 2 }  \sqrt{ - g } \Big[ - \eps g^{00} \delta g_{00}  +  p g^{11} \delta g_{11}  +  p g^{ i j } \delta g_{ i j } \Big]   \\
	&=&    \frac{ 1 }{ 2 }  \sqrt{ - g } \Big[ - \eps \Bg ^{00} \delta \Bg _{00}  +  p \Bg ^{11} \delta \Bg _{11}  +  p \chi^{ - 2 } \Fg ^{ i j } \delta \big( \chi ^2 \Fg _{ i j } \big)  \Big]  \nonumber \\
\ \\
	&=&    \frac{ 1 }{ 2 }  \sqrt{ - g } \Big[ - \eps \Bg ^{00} \delta \Bg _{00}  +  p \Bg ^{11} \delta \Bg _{11}  +  ( d - 2 ) p \chi^{ - 2 }  \delta \big( \chi ^2 \big)  \Big]  \nonumber \\
\ \\
	&=&     \frac{ 1 }{ 2 }  \sqrt{ - g } \bigg[ - \eps \Bg ^{00} \delta \Bg _{00}  +  p \Bg ^{11}  \bigg( \frac{ \del \Bg _{11} }{ \del m } \delta m  +  \frac{ \del \Bg _{11} }{ \del \chi } \delta \chi  +  \frac{ \del \Bg _{11} }{ \del \del_1 \chi }  \delta \del_1 \chi  \bigg)  \nonumber \\
		&&  +  2 ( d - 2 ) p \chi ^{ - 1 } \delta \chi  \bigg]  \\
	&=&    \frac{ 1 }{ 2 }  \sqrt{ - g } \bigg[ - \eps \Bg ^{00} \delta \Bg _{00}  +  p \Bg ^{11}  \bigg( \frac{ \del \Bg _{11} }{ \del m } \delta m  +  \frac{ \del \Bg _{11} }{ \del \chi } \delta \chi  \bigg)  +  2 ( d - 2 ) p \chi ^{ - 1 } \delta \chi  \bigg]  \nonumber \\
		&&    -   \del_1  \bigg( \frac{ 1 }{ 2 } \sqrt{ - g } p \Bg ^{11} \frac{ \del \Bg _{11} }{ \del \del_1 \chi } \bigg) \delta \chi  +  \del_1 \bigg( \frac{ 1 }{ 2 } \sqrt{ - g } p \Bg ^{11} \frac{ \del \Bg _{11} }{ \del \del_1 \chi }  \delta \chi  \bigg)  \\
	&=&     \frac{ 1 }{ 2 }  \sqrt{ - g } \bigg[ - \eps \Bg ^{00} \delta \Bg _{00}  +  p \Bg ^{11}  \bigg( \frac{ \del \Bg _{11} }{ \del m } \delta m  +  \frac{ \del \Bg _{11} }{ \del \chi } \delta \chi  \bigg)  +  2 ( d - 2 ) p \chi ^{ - 1 } \delta \chi  \bigg]  \nonumber \\
		&&    -  \del_1  \bigg( \frac{ \sqrt{ - g } p }{ \del_1 \chi }  \bigg) \delta \chi  +  \del_1 \bigg( \frac{ \sqrt{ - g } p }{ \del_1 \chi }  \delta \chi  \bigg)  \\
	&=&     \frac{ 1 }{ 2 }  \sqrt{ - g } \bigg[ - \eps \Bg ^{00} \delta \Bg _{00}  +  p \Bg ^{11}  \bigg( \frac{ \del \Bg _{11} }{ \del m } \delta m  +  \frac{ \del \Bg _{11} }{ \del \chi } \delta \chi  \bigg)  +  2 ( d - 2 ) p \chi ^{ - 1 } \delta \chi  \bigg]  \nonumber \\
		&&    -  \del_1  \bigg( \frac{ \sqrt{ - g } p }{ \del_1 \chi }  \bigg) \delta \chi  +  \del_1 \bigg( - t_0 \frac{ n_1 }{ \del_1 \chi } \sqrt{ \gamma } p  \delta \chi  \bigg)  
\end{eqnarray}

The variation of the total entropy $\delta S$ of the system that satisfies the field equations is 
\begin{eqnarray}
\delta S  +  \lam_0 \delta N   &=&   \oint _{ \del \mathcal{V} } \d ^{ d - 2 } x \frac{ u^0 }{ T } \bigg( - \frac{ 1 }{ \kappa ^2 } \D _{ \del_1 m } \sqrt{ - g } G^0 _{ \ \  0 }  \cdot  \delta m  -  t_0 \frac{ n_1 }{ \del_1 \chi } \sqrt{ \gamma } p  \delta \chi  \bigg)  \nonumber \\
	&\sseq&   \oint _{ \del \mathcal{V} } \d ^{ d - 2 } x \frac{ u^0 }{ T } ( - t_0 ) \frac{ n_1 }{ \del_1 \chi } \bigg( \frac{ \sqrt{ \Fg } }{ A_{ \mathrm F } } \delta m  +  p \sqrt{ \gamma }  \delta \chi  \bigg)  \\
	&\sseq&   \bigg(  \frac{ n_1 }{ T \del_1 \chi } \delta m  +  \frac{ n_1 }{ T \del_1 \chi } A_{ \mathrm F } \chi ^{ d - 2 } p \delta \chi  \bigg) \bigg| _{ \del \mathcal{V} }  \\
	&=:&  \frac{1}{ T_{\rm global} } \delta m |_{ \del \mathcal{V} }  +  \frac{ A_{ \mathrm F } \chi ^{ d - 2 } }{ T_{\rm global} } P_{\rm global} \delta \chi |_{ \del \mathcal{V} } .  
\end{eqnarray}
The globally defied temperature $T_{\rm global}$ and pressure $P_{\rm global}$ are  
\begin{eqnarray}
	T_{\rm global}  &=&  \frac{ \del_1 \chi }{ n_1 } T  \bigg| _{ \del \mathcal{V} } \\
	P_{\rm global}  &=&  p | _{ \del \mathcal{V} } . 
\end{eqnarray}

For example, in general relativity without the cosmological constatn in $4$-dimensional spacetime, $a_{ ( 0 ) } = 0 , \ a_{ ( 1 ) } = 1 , a_{ ( 2 ) } = \cdots =  0$, $d = 4$, $A_{ \mathrm F } = 4 \pi$, and $\kappa ^2 = 8 \pi$, 
\begin{eqnarray}
	g_{11}  &\sseq&  ( \del_1 \chi ) ^2 \bigg( 1  -  \frac{ 2 m  }{ \chi }  \bigg) ^{ - 1 } . 
\end{eqnarray}
The global temperature and pressure in general relativity are 
\begin{eqnarray}
	T_{\rm global} ^{\rm GR}  &=&   \Bigg[  \bigg( 1  -  \frac{ 2 m  }{ \chi }  \bigg) ^{ 1 / 2 }  T  \Bigg] \Bigg| _{ \del \mathcal{V} }  \\
	P_{\rm global} ^{\rm GR}  &=&   p |_{ \del \mathcal{V} } . 
\end{eqnarray}

\section{Constraints on the Lagrangian} \label{Constraint Lagrangian}
In this section, we determine the form of the Lagrangian supposing following two assumptions: 
\begin{enumerate}
\item  The variation of the entropy $\delta S$ is represented by a linear combination of the variations of a conserved energy $\delta \t M$, the particle number $\delta \t N$, and the size of the system $\delta \t \chi$ in the kinematical geometry. \label{a1}
\item  The uniform rescaling of the kinematical volume element with the thermodynamical geometry fixed, namely 
	\begin{eqnarray}
	&&  \eta \rightarrow \eta + a  , \ \ \ \ \ \   a = {\rm const.} \\
	&&  g_{ \mu \nu }  \rightarrow  g_{ \mu \nu } 
	\end{eqnarray}
does not affect the dynamics of the fields in vacuum resion. \label{a2}
\end{enumerate}

In the previous section, we saw that the Lovelock theory satisfies the assumption \ref{a1}. However, the assumption \ref{a2} is not satisfied, due to the dependence of the uniform rescaling of Lovelock's Lagrangian $L_{ ( n ) }$ on their order $n$, i.e., 
\begin{eqnarray}
L_{ ( n ) }  \rightarrow  e^{ - 2 n a } L_{ ( n ) } . 
\end{eqnarray}
Therefore, the gravitational part of the Lagrangian should be one of $n$-th order Lovelock's Lagrangians. In this work, we adopt the $1$st order, namely the Einstein-Hilbert Lagrangian $R$. 

Let us consider the Lagrangian including $A_{ \mu }$, $\eta$, and $\varphi$. For satisfying the assumption \ref{a2}, the Lagrangian should be written in the form 
\begin{eqnarray}
\mathcal{L}  =  \frac{ 1 }{ \kappa^2 } e^{ k \eta } \mathcal{K} [ g^{ \mu \nu } , A_{ \mu } , \del_{ \mu } \eta , \varphi  ] , 
\end{eqnarray}
so that the rescaling of $\eps$ and $n$ 
\begin{eqnarray}
\sqrt{ - g } \eps  &=&  - \frac{ 1 }{ \kappa^2 } u^{ \rho } u^{ \sig } e^{ - \eta } \D _{ g^{ \rho \sig } } [ \sqrt{ - g } e^{ k \eta } \mathcal{K} ] \\
	&\rightarrow&  - \frac{ e^{ ( k - 1 ) a } }{ \kappa^2 } u^{ \rho } u^{ \sig } e^{ - \eta } \D _{ g^{ \rho \sig } } [ \sqrt{ - g } e^{ k \eta } \mathcal{K} ] \\
\sqrt{ - g } n  &=&  \frac{ 1 }{ q \kappa^2 } u_{ \rho } \D _{ A_{ \rho } } [ \sqrt{ - g } e^{ k \eta } \mathcal{K} ] \\
	&\rightarrow&  \frac{ e^{ k a } }{ q \kappa^2 } u_{ \rho } \D _{ A_{ \rho } } [ \sqrt{ - g } e^{ k \eta } \mathcal{K} ]
\end{eqnarray}
can be canceled by re-defining the gravitational constant $\kappa^2$ and electrical charge density $q$, such as $\kappa^2 =  \kappa^2 _0 e^{ ( k - 1 ) \eta_0 }$, $q  =  q _0 e^{ \eta _0 }$, and $\eta _0$ is the value of $\eta$ at an arbitrary point.

Imposing the $U(1)$ gauge invariance (\ref{gaugephi}) and (\ref{gaugeA}), Noether's identity (\ref{id5}) shows that the conservation law of $N$ demands the field $\eta$ to be electrically neutral. The following Lagrangian density satisfies all the demands: 
\begin{eqnarray}
\mathcal{L}  =  \frac{ 1 }{ \kappa ^2 } e^{ ( d - 2 ) \eta }  \bigg(  R  -  \frac{ 1 }{ 4 } F_{ \mu \nu } F^{ \mu \nu }  +  \omega \del_{ \mu } \eta \del^{ \mu } \eta  -  D_{ \mu } \varphi D^{ \mu } \varphi  \bigg) , \label{Lagtherm}
\end{eqnarray}
where $D_{ \mu } \varphi  :=  ( \del_{ \mu }  -  i e_{ \varphi } A_{ \mu } ) \varphi$ is the gauge covariant derivative. 

Under the conformal transformation (\ref{defGDtilde}), the scalar curvature $R$ transforms into 
\begin{eqnarray}
R  =  e^{ 2 \eta }  \Big[  \t R  + 2 ( d - 1 ) \t \nabla ^2 \eta  -  ( d - 2 ) ( d - 1 ) \del_{ \mu } \eta \t \del^{ \mu } \eta  \Big] , 
\end{eqnarray}
and the Lagrangian density in thermodynamical frame (\ref{Lagtherm}) into kinematical frame as 
\begin{eqnarray}
\sqrt{ - g }  \mathcal{L}  &=&  \frac{ 1 }{ \kappa^2 } \sqrt{ - \t g }  \bigg[ \t R  + 2 ( d - 1 ) \t \nabla ^2 \eta  -  ( d - 2 ) ( d - 1 ) \del_{ \mu } \eta \t \del^{ \mu } \eta   \nonumber \\
	&&  -  \frac{ 1 }{ 4 } e^{ 2 \eta }  F_{ \mu \nu } \t F^{ \mu \nu }  +  \omega \t g^{ \mu \nu } \del_{ \mu } \eta  \del_{ \nu } \eta  -  \t g^{ \mu \nu } D_{ \mu } \varphi D_{ \nu } \varphi   \bigg]  \\
	&=&  \frac{ 1 }{ \kappa^2 } \sqrt{ - \t g } \bigg\{  \t R  -  \frac{ 1 }{ 4 } e^{ 2 \eta }  F_{ \mu \nu } \t F^{ \mu \nu }  +  \big[ \omega  -  ( d - 2 ) ( d - 1 )  \big]  \del_{ \mu } \eta \t \del^{ \mu } \eta  -  D_{ \mu } \varphi \t D^{ \mu } \varphi  \bigg\}  \nonumber \\
	&&  +  \del_{ \mu } \Big[  2 ( d - 1 ) \sqrt{ - \t g } \t \del^{ \mu } \eta \Big] . 
\end{eqnarray}
Omitting the total derivative, the Lagrangian density in kinematical frame $\sqrt{ - \t g } \mathcal{\t L}$ satisfying the assumption \ref{a1} and \ref{a2} is written by 
\begin{eqnarray}
\sqrt{ - \t g } \mathcal{ \t L }  =  \frac{ 1 }{ \kappa^2 } \sqrt{ - \t g } \bigg\{  \t R  -  \frac{ 1 }{ 4 } e^{ 2 \eta } F_{ \mu \nu } \t F^{ \mu \nu }  +  \big[ \omega  -  ( d - 2 ) ( d - 1 )  \big]  \del_{ \mu } \eta \t \del^{ \mu } \eta  -  D_{ \mu } \varphi \t D^{ \mu } \varphi   \bigg\} .  \nonumber \\
\end{eqnarray}

\section{Conclusion and discussion} \label{Conclusion}

We calculate the first order variation of the entropy of a self-interacting system using only the first law of thermodynamics (\ref{FL}), the Gibbs-Duhem relation (\ref{GDR}), and Noether's theorem for the invariance of the coordinate and $U( 1 )$ gauge transformations. We find that as long as there exists a Lagrangian density $\mathcal{L}$ and the ordinary constraint relation 
\begin{eqnarray}
\sqrt{ - g } \eps  =  - u^{ \mu} u^{ \nu } \D _{ g^{ \mu \nu } } \sqrt{ - g } \mathcal{L}  \label{constA}
\end{eqnarray}
is imposed, every stationary state satisfying the following ordinary field equations for a charged perfect fluid 
\begin{eqnarray}
&&  \delta \eps  =  T \delta s  +  \mu \delta n  \label{FL2} \\
&&  T s  +  \mu n  =  \eps  +  p  \label{GDR2} \\
&&  \D _{ g^{ \mu \nu } } \sqrt{ - g } \mathcal{L}  =  \sqrt{ - g } \Big[  \eps u_{ \mu } u_{ \nu } + p ( u_{ \mu } u_{ \nu }  +  g_{ \mu \nu } )  \Big]  \label{feG} \\
&&  \D _{ A_{ \mu } } \sqrt{ - g } \mathcal{L}  =  \sqrt{ - g } u^{ \mu } q n \label{feA} \\
&&  \D _{ \varphi } \sqrt{ - g } \mathcal{L}  =  0 \label{fephi}
\end{eqnarray}
maximizes the total entropy under appropriate boundary conditions, no matter what the form of the Lagrangian density $\mathcal{L}$ that depends on $g^{ \mu \nu }$, $A_{ \mu }$, $\varphi$, and $u^{ \mu }$. Therefore, the compatibility between ordinary field equations (\ref{FL2}) - (\ref{fephi}) and the maximum entropy principle requires a physical necessity for the constraint relation (\ref{constA}) to hold.

Further, enlarging the region of the constraint relation to 
\begin{eqnarray}
\sqrt{ - g } \eps  =  - e^{ - \eta } u^{ \mu } u^{ \nu } \D _{ g^{ \mu \nu } } \sqrt{ - g } \mathcal{L} , \label{constB}
\end{eqnarray}
and varying the additional field $\eta$ independently, we find that the modified field equations (\ref{ELeqG}) - (\ref{ELeqphi}) maximize the entropy of the stationary states, instead a force-free fluid do not flow along the geodesic orbits for $g_{ \mu \nu }$ so that another geometry $\t g_{ \mu \nu }$, the kinematical geometry, should be introduced. The kinematical geometry $\t g_{ \mu \nu }$ and the originally introduced thermodynamical geometry $g_{ \mu \nu }$ are related by the following conformal transformation 
$$\t g_{ \mu \nu }  =   e^{ 2 \eta } g_{ \mu \nu } . $$
In the kinematical geometry, the modified field equations (\ref{ELeqG}) - (\ref{ELeqphi}) transform into 
\begin{eqnarray}
&&  \delta \t \eps  =  T \delta \t s  -  p \delta \t v  +  \mu \delta \t n \label{FLtilde2} \\
&&  T \t s  +  \mu \t n  =  \t \eps  +  p \t v \label{GDRtilde2} \\
&&  \D _{ \t g ^{\mu \nu } }  \sqrt{ - \t g } \mathcal{ \t L }  =   \sqrt{ - \t g } \Big[  \t \eps \t u_{\mu } \t u_{\nu } + p \t v ( \t u_{ \mu } \t u_{ \nu }  +  \t g_{\mu \nu } )  \Big]  \\
&&  \D _{ A_{\mu } } \sqrt{ - \t g } \mathcal{ \t L }  =  \sqrt{ - \t g } \t u^{ \mu } q  \t n   \\
&&  \D _{ \eta }  \sqrt{ - \t g } \mathcal{ \t L }  =   2 ( d - 1 ) \sqrt{ - \t g } p \t v  \label{ELeqetatilde2} \\
&&  \D _{\varphi } \sqrt{ - \t g } \mathcal{ \t L}  =  0 , \label{ELeqphitilde2}
\end{eqnarray}
where $\t v = e^{ - ( d - 1 ) \eta }$ is interpreted as the (thermodynamical) volume density measured by the kinematical volume element. 

Concrete calculations show that the variation of the entropy in the Lovelock theory of gravity is expressed by a linear combination of the variations of the generalized Kodama energy and the size of the system. Further, demanding that uniform rescaling the volume density $\t v$ with the thermodynamical geometry $g_{ \mu \nu }$ fixed should not affect the dynamics of the fields in vacuum region, the appropriate form of Lagrangian density is determined and expressed in the thermodynamical frame as 
\begin{eqnarray}
\mathcal{L}  =  \frac{ 1 }{ \kappa ^2 } e^{ ( d - 2 ) \eta }  \bigg(  R  -  \frac{ 1 }{ 4 } F_{ \mu \nu } F^{ \mu \nu }  +  \omega \del_{ \mu } \eta \del^{ \mu } \eta  -  D_{ \mu } \varphi D^{ \mu } \varphi  \bigg) , \label{LagTherm}
\end{eqnarray}
and in the kinematical frame as 
\begin{eqnarray}
\mathcal{ \t L }  =  \frac{ 1 }{ \kappa^2 } \bigg\{  \t R  -  \frac{ 1 }{ 4 } e^{ 2 \eta } F_{ \mu \nu } \t F^{ \mu \nu }  +  \big[ \omega  -  ( d - 2 ) ( d - 1 )  \big]  \del_{ \mu } \eta \t \del^{ \mu } \eta  -  D_{ \mu } \varphi \t D^{ \mu } \varphi   \bigg\} .  \label{LagKinem} \nonumber \\
\end{eqnarray}
Including an anti-symmetric tensor $B_{ \mu \nu }$ is straightforward. It is strange and interesting that these Lagrangians (\ref{LagTherm}) and (\ref{LagKinem}) for $\omega = ( d - 2 ) ^2$ are analogous to the low energy effective theories of bosonic string \cite{STRING} for the critical dimension, namely $d = 26$, in the string frame and Einstein frame, respectively, and the field $\eta$ corresponds to the dilaton field $\Phi$, even though there is no direct relationship between the two.  In string theory, the dilaton action is obtained by requiring the quantum conformal anomaly of the Polyakov action on world sheet to vanish. This analogy implies that the maximum entropy principle in a spacetime be related to the quantum conformal symmetry of the Polyakov action of wold sheet embedded into the spacetime as the background geometry.  

The field equations (\ref{FLtilde2}) - (\ref{ELeqphitilde2}) are strange but have some desirable eatures. The existence of a source term proportional to Gibbs's free energy density $p \t v$ in Eq. (\ref{ELeqetatilde2}) leads some interesting properties for application to cosmology and black hole physics. First, in application to cosmology, since the source term of Eq. (\ref{ELeqetatilde2}) can lead acceleration of expansion of universe, it may explain the data of cosmological observations without introducing dark energy, the cosmological constant, nor inhomogeneity of universe. Second, in application to black hole physics, the non zero source term in Eq. (\ref{ELeqetatilde2}) leads an irremovable scalar hair around a black hole, so that the uniqueness theorem, or so-called no-hair theorem \cite{Israel1967, Carter1971, Robinson1975} prevents the black hole from being spherically symmetric and static, i.e., a black hole has no spherically symmetric and static state. This can be rephrased as follows: in the theories that respect the maximal entropy principle, a spherically symmetric and static black hole, for example a Schwarzchild black hole, has a kind of instability to thermal fluctuations.

\section*{Acknowledgements}
The author would like to thank all who supported this research and gave me fruitful discussions.

\appendix

\section{The homogeneity of the extensive quantities} \label{Homogeneity}

Consider a $( d - 1 )$-dimensional region $\Omega$ whose length scale is sufficiently small than that of the field configuration. The thermodynamical volume of $\Omega$, measured by $g_{ \mu \nu }$, is $\Delta V$. Since the entropy $\Delta S$, energy $\Delta E$, and particle number $\Delta N$ are extensive quantities, they are represented by the products of their densities and $\Delta V$, namely, $\Delta S = s \Delta V$, $\Delta E = \eps \Delta V$, and $\Delta N = n \Delta V$. The entropy $\Delta S$ is a function of $\Delta E$, $\Delta N$, and $\Delta V$, i.e., 
\begin{eqnarray}
\Delta S  &=&  \Delta S ( \Delta E , \Delta N , \Delta V )  \\
	&=&  s ( \Delta E / \Delta V , \Delta N / \Delta V ) \, \Delta V . \label{DeltaS}
\end{eqnarray}
Thus, the function $s ( \eps , n )$ is represented by 
\begin{eqnarray}
s ( \eps , n )  =  \frac{ 1 }{ \Delta V }  \Delta S ( \eps \Delta V  , n \Delta V , \Delta V ) . 
\end{eqnarray}
From the definitions of $T$ and $\mu$, namely Eq. (\ref{defT}) and (\ref{defMu}), 
\begin{eqnarray}
\frac{ \del \Delta S }{ \del \Delta E }  =  \frac{ \del s }{ \del \eps }  =  \frac{ 1 }{ T }  \\
\frac{ \del \Delta S }{ \del \Delta N }  =  \frac{ \del s }{ \del n }  =  - \frac{ \mu }{ T } . 
\end{eqnarray}
The equation (\ref{DeltaS}) shows that the function $\Delta S ( \Delta E , \Delta N , \Delta V )$ is a homogeneous function of degree $1$. 
\begin{eqnarray}
\forall a \in \mathbb{R}, \ \ \ \ \ \  a \Delta S ( \Delta E , \Delta N , \Delta V )  =  \Delta S ( a \Delta E , a \Delta N , a \Delta V ) . \label{homogenS}
\end{eqnarray}
The entropy density $s( \eps , n )$ is written by 
\begin{eqnarray}
s( \eps , n )  =  \Delta S ( \eps , n , 1 ) . 
\end{eqnarray}

Conversely, it is necessary for constructing $s( \eps , n )$ that the function $\Delta S ( \Delta E , \Delta N , \Delta V )$ is homogeneous of degree $1$. Therefore, the homogeneity of $\Delta S ( \Delta E , \Delta N , \Delta V )$ is a necessary and sufficient condition of the extensiveness of $s$, $\eps$, and $n$.

Differentiating Eq. (\ref{homogenS}) with respect to $a$ and substituting $a = 1$, one obtains following relation: 
\begin{eqnarray}
\Delta S  &=&  \frac{ \del \Delta S }{ \del \Delta E } \Delta E  + \frac{ \del \Delta S }{ \del \Delta N } \Delta N  +  \frac{ \del \Delta S }{ \del \Delta V } \Delta V  \\
	&=&  \frac{ 1 }{ T } \Delta E  -  \frac{ \mu }{ T } \Delta N  +  \frac{ \del \Delta S }{ \del \Delta V } \Delta V  . \label{preFL}
\end{eqnarray}
Dividing by $\Delta V$ and multiplying $T$, 
\begin{eqnarray}
T s + \mu n  =  \eps  +  T \frac{ \del \Delta S }{ \del \Delta V } . \label{preGDR}
\end{eqnarray}
Comparing Eq. (\ref{preGDR}) with the definition of $p$, namely Eq. (\ref{defp}), 
\begin{eqnarray}
\frac{ \del \Delta S }{ \del \Delta V }  =  \frac{ p }{ T }  . \label{p over T}
\end{eqnarray}
From Eq. (\ref{preFL}) and (\ref{p over T}), 
\begin{eqnarray}
\Delta S  =   \frac{ 1 }{ T } \Delta E  -  \frac{ \mu }{ T } \Delta N  +  \frac{ p }{ T } \Delta V . 
\end{eqnarray}
This is the first law of thermodynamics. Therefore, the definition of $p$, Eq. (\ref{defp}) is equivalent to the first law.

\section{Noether's second theorem}
In this section, Noether's theorem \cite{Noether1918} for arbitrarily high order derivative is presented. 
\subsection{Coordinate transformation}
Consider the infinitesimal coordinate transformation 
\begin{eqnarray}
y^{\mu } = x^{\mu } + \td x ^{\mu } . 
\end{eqnarray}
We use the symbol $\td$ for the difference between the values of fields at the same wold point, and the symbol $\delta $ for that of the values at the same two points whose values of the coordinates coincide, i.e.,  
\begin{eqnarray}
\td f (x)  &:=&  f ^{\prime } (y)  -  f (x) \\
	&=&  \delta f (x) + \big( \del_{\mu } f (x) \big) \td x ^{\mu }  
\end{eqnarray}
and 
\begin{eqnarray}
\delta f (x)  :=  f ^{\prime } (x)  -  f (x) . 
\end{eqnarray}
The operator $\del_{ \mu } $ commutes with $\delta $, whereas does not with $\td $: 
\begin{eqnarray}
\del_{\mu } \delta f - \delta \del_{\mu } f  &=&  0 \\
\del_{\mu } \td f - \td \del_{\mu } f  &=&  (\del_{\nu } f ) \del_{\mu } \td x^{\nu } . 
\end{eqnarray}
The commutator of $\del_{ \mu } $ and $\td $ is 
\begin{eqnarray}
\big[  \del_{ \mu } , \td  \big]  =  ( \del_{ \mu } \td x^{ \nu } ) \del_{ \nu } . 
\end{eqnarray}
The commutator of $\del_{ \mu_1 } \cdots \del_{ \mu_k } $ and $\td $ is calculated as 
\begin{eqnarray}
\big[  \del_{ \mu_1 } \cdots \del_{ \mu_k } , \td  \big] f  &=&   \sum_{ l = 1 } ^{k}  \del_{ \mu_1 } \cdots \del_{ \mu_{ l - 1 } } \big[ \del_{ \mu_l } , \td \big] \del_{ \mu_{ l + 1 } } \cdots \del_{ \mu_k }  f  \\
	&=&  \sum_{ l = 1 } ^{k}  \del_{ \mu_1 } \cdots \del_{ \mu_{ l - 1 } }  \Big[ ( \del_{ \mu_l } \td x^{ \nu } ) \del_{ \nu }   \del_{ \mu_{ l + 1 } } \cdots \del_{ \mu_k }  f  \Big] . \label{commutdels}
\end{eqnarray} 
The infinitesimal coordinate transformation of scalar $\varphi $, covariant vector $A_{ \mu } $, contravariant vector $u^{ \mu }$, $2$-th order symmetric cotravariant tensor $g^{ \mu \nu } $, and $2$-th order anti-symmetric covariant tensor $B_{ \mu \nu }$ at the same world point are
\begin{eqnarray}
\td \varphi  &=&  0 \label{deltaphi} \\
\td A _{\mu }  &=&  - A_{\rho } ( \del_{\mu } \td x^{\rho } ) \label{deltaA} \\
\td u ^{\mu }  &=&  u^{\rho } ( \del_{\rho } \td x^{\mu } ) \label{deltaU} \\
\td g ^{\mu \nu }  &=&  g^{\rho \nu } ( \del_{\rho } \td x^{\mu } )  +  g^{\mu \rho } ( \del_{\rho } \td x^{\nu } ) \label{deltaG} \\
\td B _{\mu \nu }  &=&  - B _{\rho \nu } ( \del_{\mu } \td x^{\rho } )  -  B _{\mu \rho } ( \del_{\nu } \td x^{\rho } ) . \label{deltaB}
\end{eqnarray}
The transformation of their $k$-th order derivatives are calculated by differentiating Eqs. (\ref{deltaphi}) - (\ref{deltaB}) and using the comutator (\ref{commutdels}): 
\begin{eqnarray}
&& \td \del_{\mu_1 } \cdots \del_{\mu_k } \varphi  \nonumber \\
	&=&  - \sum_{ l = 1 } ^k \del_{ \mu_1 } \cdots \del_{ \mu_{ l - 1 } } \Big[ ( \del_{ \mu_l } \td x ^{ \nu } ) \del_{ \nu } \del_{ \mu_{ l + 1 } } \cdots \del_{ \mu_k } \varphi  \Big]  \label{deltaphidels} \\
	&=&  - \sum_{ l = 1 } ^k  \big( \del_{ \nu } \del_{ \mu_1 } \cdots \del_{ l - 1 } \del_{ l + 1 } \cdots \del_{ \mu_k } \varphi  \big)  \del_{ \mu_l } \td x ^{ \nu }  + \cdots  \\
	&=&  - \sum_{ l = 1 } ^k \big( \del_{ \nu } \del_{ \mu_1 } \cdots \del_{ l - 1 } \del_{ l + 1 } \cdots \del_{ \mu_{ k } } \varphi  \big) \delta ^{ \mu } _{ \mu_l } \del_{ \mu } \td x ^{ \nu }  +  \cdots , 
\end{eqnarray}
\begin{eqnarray}
&& \td \del_{\mu_1 } \cdots \del_{\mu_k } A_{\alpha }  \nonumber \\
	&=&  - \sum _{j = 0} ^{k} 
	\bigg(
    \begin{array}{c}
      k \\
      j 
    \end{array}
	\bigg)
( \del_{\mu_{j + 1} } \cdots \del_{\mu_k } A_{\gamma } ) \del_{\mu_1 } \cdots \del_{\mu_j } \del_{\alpha } \td x^{\gamma } \nonumber \\
&&  - \sum_{ l = 1 } ^k \del_{ \mu_1 } \cdots \del_{ \mu_{ l - 1 } } \Big[ ( \del_{ \mu_l } \td x ^{ \nu } ) \del_{ \nu } \del_{ \mu_{ l + 1 } } \cdots \del_{ \mu_k } A_{ \alpha }  \Big] \\ 
	&=&  - \Big[ \delta ^{\mu } _{\alpha } ( \del_{\mu_1 } \cdots \del_{\mu_k } A_{\nu } )  +  \sum_{ l = 1 } ^k ( \del_{\nu } \del_{ \mu_1 } \cdots \del_{ l - 1 } \del_{ l + 1 } \cdots \del_{ \mu_{ k } } A_{\alpha } ) \delta ^{ \mu } _{ \mu_l }  \Big] \del_{\mu } \td x^{\nu }  \nonumber \\
	 && +  \cdots , 
\end{eqnarray}
\begin{eqnarray}
&& \td \del_{\mu_1 } \cdots \del_{\mu_k } u^{\alpha }  \nonumber \\
	&=&   \sum _{j = 0} ^{k} 
	\bigg(
    \begin{array}{c}
      k \\
      j 
    \end{array}
	\bigg)
( \del_{\mu_{j + 1} } \cdots \del_{\mu_k } u^{\gamma } ) \del_{\mu_1 } \cdots \del_{\mu_j } \del_{\gamma } \td x^{\alpha } \nonumber \\
&&   - \sum_{ l = 1 } ^k \del_{ \mu_1 } \cdots \del_{ \mu_{ l - 1 } } \Big[ ( \del_{ \mu_l } \td x ^{ \nu } ) \del_{ \nu } \del_{ \mu_{ l + 1 } } \cdots \del_{ \mu_k } u^{ \alpha }  \Big]  \\
	&=&   \Big[ \delta _{\nu } ^{\alpha } ( \del_{\mu_1 } \cdots \del_{\mu_k } u^{\mu } )  - \sum_{ l = 1 } ^k ( \del_{\nu } \del_{ \mu_1 } \cdots \del_{ l - 1 } \del_{ l + 1 } \cdots \del_{ \mu_{ k } } u^{\alpha } ) \delta ^{ \mu } _{ \mu_l } \Big] \del_{\mu } \td x^{\nu }  \nonumber \\
	 && +  \cdots , 
\end{eqnarray}
\begin{eqnarray}
&& \td \del_{\mu_1 } \cdots \del_{\mu_k } g^{\alpha \beta }  \nonumber \\
	&=&    \sum _{j = 0} ^{k} 
	\bigg(
    \begin{array}{c}
      k \\
      j 
    \end{array}
	\bigg)
( \del_{\mu_{j + 1} } \cdots \del_{\mu_k } g^{ \gamma \beta } ) \del_{\mu_1 } \cdots \del_{\mu_j } \del_{\gamma } \td x^{\alpha } \nonumber \\
&& + \sum _{j = 0} ^{k} 
	\bigg(
    \begin{array}{c}
      k \\
      j 
    \end{array}
	\bigg)
( \del_{\mu_{j + 1} } \cdots \del_{\mu_k } g^{\alpha \gamma } ) \del_{\mu_1 } \cdots \del_{\mu_j } \del_{\gamma } \td x^{\beta } \nonumber \\
&&    - \sum_{ l = 1 } ^k \del_{ \mu_1 } \cdots \del_{ \mu_{ l - 1 } } \Big[ ( \del_{ \mu_l } \td x ^{ \nu } ) \del_{ \nu } \del_{ \mu_{ l + 1 } } \cdots \del_{ \mu_k } g^{ \alpha \beta }  \Big]  \\
	&=&  \Big[ \delta ^{\alpha } _{\nu } ( \del_{\mu_1 } \cdots \del_{\mu_k } g^{ \mu \beta } )  +  \delta ^{\beta } _{\nu } ( \del_{\mu_1 } \cdots \del_{\mu_k } g^{\alpha \mu } )   -  \sum_{ l = 1 } ^k ( \del_{\nu } \del_{ \mu_1 } \cdots \del_{ l - 1 } \del_{ l + 1 } \cdots \del_{ \mu_{ k } } g^{ \alpha \beta } ) \delta ^{ \mu } _{ \mu_l } \Big]  \del_{ \mu } \td x^{ \nu } \nonumber \\
	&& + \cdots . 
\end{eqnarray}
\begin{eqnarray}
&& \td \del_{\mu_1 } \cdots \del_{\mu_k } B_{ \alpha \beta }  \nonumber \\
	&=&    - \sum _{j = 0} ^{k} 
	\bigg(
    \begin{array}{c}
      k \\
      j 
    \end{array}
	\bigg)
( \del_{ \mu_{ j + 1 } } \cdots \del_{ \mu_k } B_{ \gamma \beta } ) \del_{\mu_1 } \cdots \del_{\mu_j } \del_{ \alpha } \td x^{ \gamma } \nonumber \\
&& - \sum _{j = 0} ^{k} 
	\bigg(
    \begin{array}{c}
      k \\
      j 
    \end{array}
	\bigg)
( \del_{\mu_{j + 1} } \cdots \del_{\mu_k } B_{\alpha \gamma } ) \del_{\mu_1 } \cdots \del_{\mu_j } \del_{ \beta } \td x^{ \gamma } \nonumber \\
&&    - \sum_{ l = 1 } ^k \del_{ \mu_1 } \cdots \del_{ \mu_{ l - 1 } } \Big[ ( \del_{ \mu_l } \td x ^{ \nu } ) \del_{ \nu } \del_{ \mu_{ l + 1 } } \cdots \del_{ \mu_k } B_{ \alpha \beta }  \Big]  \\
	&=&  - \Big[ \delta ^{ \mu } _{ \alpha } ( \del_{\mu_1 } \cdots \del_{\mu_k } B_{ \nu \beta } )  +  \delta ^{ \mu } _{ \beta } ( \del_{\mu_1 } \cdots \del_{\mu_k } B_{ \alpha \nu } )   +  \sum_{ l = 1 } ^k ( \del_{\nu } \del_{ \mu_1 } \cdots \del_{ l - 1 } \del_{ l + 1 } \cdots \del_{ \mu_{ k } } B_{ \alpha \beta } ) \delta ^{ \mu } _{ \mu_l } \Big]  \del_{ \mu } \td x^{ \nu } \nonumber \\
	&& + \cdots , \label{deltaBdels1}
\end{eqnarray}
where the last line of each equation is the term which is proportional to the least derivative, namely the first derivative, of $\td x^{ \nu }$. 

The infinitesimal transformation of $\d ^n x $ is 
\begin{eqnarray}
\d ^n y  =  \big( 1 + \del_{\mu } \td x ^{\mu } \big) \d ^n x . 
\end{eqnarray}

Consider an arbitrary scalar function $\mathcal{F}$.  The infinitesimal transformation of the integral of $\sqrt{ - g } \mathcal{F}$ over $\mathcal{M}$ is arranged as 
\begin{eqnarray}
	&&  \int _{ \mathcal{M} } \d^n y \sqrt{ - g^{\prime } } \mathcal{F} ^{\prime } (y)  -  \int _{ \mathcal{M} } \d^n x \sqrt{ - g } \mathcal{F} (x)  \\
	&=&  \int _{\mathcal{M}} \d^n x \Big[ \td \big( \sqrt{ - g } \mathcal{F} \big) + \sqrt{ - g } \mathcal{F} \big( \del_{\mu } \td x^{\mu }  \big) \Big]  \\
	&=&  \int _{\mathcal{M}} \d^n x \bigg[ \frac{\del (\sqrt{ - g } \mathcal{F} ) }{\del f } \td f + \frac{\del ( \sqrt{ - g } \mathcal{F} ) }{\del (\del_{\mu_1 } f )} \td \del_{\mu_1 } f + \cdots + \sqrt{ - g } \mathcal{F} \del_{\mu } \td x^{\mu  }  \bigg]  \\
	&=&  \int _{\mathcal{M}} \d^n x \bigg[ \frac{\del ( \sqrt{ - g } \mathcal{F} ) }{\del f } \delta f + \frac{\del ( \sqrt{ - g } \mathcal{F} ) }{\del (\del_{\mu_1 } f )} \delta \del_{\mu_1 } f + \cdots + \del_{\mu } \big( \sqrt{ - g } \mathcal{F} \td x^{\mu  } \big)  \bigg] \\
	&=&  \int _{\mathcal{M}} \d^n x \bigg[ \D_f  \sqrt{ - g } \mathcal{F} \cdot \delta f \nonumber \\
	&& +  \del_{\mu } \bigg( \sum_{k= 0} ^{\infty } \D_{\del_{\mu } \del_{\mu_1 } \cdots \del_{\mu_k } f }  \sqrt{ - g } \mathcal{F} \cdot \delta \del_{\mu_1 } \cdots \del_{\mu_k } f  +  \sqrt{ - g } \mathcal{F} \td x^{\mu }  \bigg)   \bigg]  \\
	&=&  \int _{\mathcal{M}} \d^n x \bigg[ \D_f  \sqrt{ - g } \mathcal{F} \cdot \delta f \nonumber \\
	&& +  \del_{\mu } \bigg( \sum_{k= 0} ^{\infty } \D_{\del_{\mu } \del_{\mu_1 } \cdots \del_{\mu_k } f }  \sqrt{ - g } \mathcal{F} \cdot \td \del_{\mu_1 } \cdots \del_{\mu_k } f  -  \sqrt{ - g } T ^{\mu } _{\ \ \nu }  [\sqrt{ - g } \mathcal{F}]  \td x^{\nu }  \bigg)   \bigg] \nonumber \\
	\\
	&=&  \int _{\mathcal{M}} \d^n x \bigg[ \D_f  \sqrt{ - g } \mathcal{F} \cdot \td f - (\del_{\mu } f ) \D_f   \sqrt{ - g } \mathcal{F} \cdot \td x^{\mu } \nonumber \\
	&& +  \del_{\mu } \bigg( \sum_{k= 0} ^{\infty } \D_{\del_{\mu } \del_{\mu_1 } \cdots \del_{\mu_k } f }  \sqrt{ - g } \mathcal{F} \cdot \td \del_{\mu_1 } \cdots \del_{\mu_k } f  -  \sqrt{ - g } T ^{\mu } _{\ \ \nu } [ \sqrt{ - g } \mathcal{F}] \td x^{\nu }  \bigg)   \bigg] , \label{deltaIntegral} \nonumber \\
\end{eqnarray}
where $T^{ \mu } _{ \ \ \nu } [ \sqrt{ - g } \mathcal{F} ]$ is defined by 
\begin{eqnarray}
\sqrt{ - g } T ^{\mu } _{\ \ \nu } [\sqrt{ - g } \mathcal{F}]  &:=&  \sum_{k= 0} ^{\infty } \D_{\del_{\mu } \del_{\mu_1 } \cdots \del_{\mu_k } f }  \sqrt{ - g } \mathcal{F} \cdot \del_{\nu } \del_{\mu_1 } \cdots \del_{\mu_k } f  -  \sqrt{ - g } \mathcal{F} \delta ^{\mu } _{\nu } .  \nonumber \\
\end{eqnarray}
When $\mathcal{F}$ is the Lagrangian of a system, $T^{ \mu } _{ \ \ \nu } [ \sqrt{ - g } \mathcal{L} ]$ is called the canonical energy-momentum affine tensor.

\subsection{Identities derived from the coordinate transformation invariance}
Because the integral of a scalar density $\sqrt{ - g } \mathcal{F}$ is invariant under any coordinate transformations, the integral (\ref{deltaIntegral}) vanishes for arbitrary $\td x^{ \mu } $. When $f = \{ g^{\alpha \beta } , B_{ \alpha \beta } , A_{\alpha } , u^{ \alpha } , \varphi \} $, substituting\footnote{We use the symmetries $g^{ \mu \nu } = g^{ \nu \mu } $, $B_{ \mu \nu }  =  - B_{ \nu \mu }$, and 
$$ \D _{ g^{ \mu \nu } } \sqrt{ - g } \mathcal{F}  =  \D _{ g^{ \nu \mu } } \sqrt{ - g } \mathcal{F} $$
$$ \D _{ B_{ \mu \nu } } \sqrt{ - g } \mathcal{F}  =  - \D _{ B_{ \nu \mu } } \sqrt{ - g } \mathcal{F} . $$ } Eqs. (\ref{deltaphi}) - (\ref{deltaB}) into Eq. (\ref{deltaIntegral}), 
\begin{eqnarray}
0  &=&   \int _{\mathcal{M}} \d^n x \bigg\{ - \Big[   \del_{\nu } \Big( 2 g^{\nu \beta } \D_{g^{\mu \beta } } \sqrt{ - g } \mathcal{F}  -  2 B_{ \mu \beta } \D _{ B_{ \nu \beta } } \sqrt{ - g } \mathcal{F}  -  A_{\mu } \D _{A_{\nu } }  \sqrt{ - g } \mathcal{F}  \nonumber \\
	&&  +  u^{ \nu } \D _{ u^{ \mu } } \sqrt{ - g } \mathcal{F}   \Big)  +  ( \del_{\mu } f ) \D_f  \sqrt{ - g } \mathcal{F}     \Big] \td x^{\mu }  \nonumber \\
	&& +  \del_{\mu } \bigg[  \Big( 2  g^{\mu \beta } \D_{g^{\nu \beta } }  \sqrt{ - g } \mathcal{F}  -  2 B_{ \nu \beta } \D _{ B_{ \mu \beta } } \sqrt{ - g } \mathcal{F}  -  A_{\nu } \D_{A_{\mu } }  \sqrt{ - g } \mathcal{F}  \nonumber \\
	&&  +  u^{ \mu } \D _{ u^{ \nu } } \sqrt{ - g } \mathcal{F}   -  \sqrt{ - g } T ^{\mu } _{\ \ \nu }  [ \sqrt{ - g } \mathcal{F} ]   \Big) \td x^{\nu }  \nonumber \\
	&& +  \sum_{k = 0} ^{\infty }  \D_{\del_{\mu } \del_{\mu_1 } \cdots \del_{\mu_k } f }  \sqrt{ - g } \mathcal{F}  \cdot  \td \del_{\mu_1 } \cdots \del_{\mu_k } f  \bigg] \bigg\} \label{deltaIntegral2a} \\
	&=&  - \int _{\mathcal{M}} \d^n x  \Big[   \del_{\nu } \Big( 2 g^{\nu \beta } \D_{g^{\mu \beta } }  \sqrt{ - g } \mathcal{F}  -  2 B_{ \mu \beta } \D _{ B_{ \nu \beta } } \sqrt{ - g } \mathcal{F}  -  A_{\mu } \D _{A_{\nu } }  \sqrt{ - g } \mathcal{F}  \nonumber \\
	&&  +  u^{ \nu } \D _{ u^{ \mu } } \sqrt{ - g } \mathcal{F}   \Big)  +  ( \del_{\mu } f ) \D_f  \sqrt{ - g } \mathcal{F}     \Big] \td x^{\mu }  \nonumber \\
	&& +  \oint _{\del \mathcal{M}} ( \d^{n - 1} x ) _{\mu } \bigg[  \Big( 2  g^{\mu \beta } \D _{g^{\nu \beta } }  \sqrt{ - g } \mathcal{F}  -  2 B_{ \nu \beta } \D _{ B_{ \mu \beta } } \sqrt{ - g } \mathcal{F}  -  A_{\nu } \D _{A_{\mu } }  \sqrt{ - g } \mathcal{F}  \nonumber \\
	&&  +  u^{ \mu } \D _{ u^{ \nu } } \sqrt{ - g } \mathcal{F}  -  \sqrt{ - g } T ^{\mu } _{\ \ \nu }  [ \sqrt{ - g } \mathcal{F} ]   \Big) \td x^{\nu }  +  \sum_{k = 0} ^{\infty }  \D _{\del_{\mu } \del_{\mu_1 } \cdots \del_{\mu_k } f }  \sqrt{ - g } \mathcal{F}  \cdot  \td \del_{\mu_1 } \cdots \del_{\mu_k } f  \bigg]  . \nonumber \\ \label{deltaIntegral2b}
\end{eqnarray}
Consider a coordinate transformation such that $\td x^{ \mu } = \del_{ \nu_1 } \td x^{ \mu } = \del_{ \nu_1 } \del_{ \nu_2 } \td x^{ \mu } = \cdots = 0$ on $\del \mathcal{M}$, in other words, the third and fourth lines of (\ref{deltaIntegral2b}) vanish. Then, the square bracket in the first and second lines of (\ref{deltaIntegral2b}) should be identically $0$, i.e., 
\begin{eqnarray}
	&&  \del_{\nu } \Big( 2 g^{ \nu \beta } \D _{ g^{ \mu \beta  } }  \sqrt{ - g } \mathcal{F}  -  2 B_{ \mu \beta } \D _{ B_{ \nu \beta } } \sqrt{ - g } \mathcal{F}  -  A_{\mu } \D _{ A_{\nu } }  \sqrt{ - g } \mathcal{F}  +  u^{ \nu } \D _{ u^{ \mu } } \sqrt{ - g } \mathcal{F}   \Big)  \nonumber \\
	&& +  ( \del_{\mu } f ) \D _f  \sqrt{ - g } \mathcal{F}    \equiv  0 .  \nonumber \\  \label{id1}
\end{eqnarray}
Introducing the symbols $E_{ \mu \nu }$, $\bar E _{ \mu \nu }$, and $F_{ \mu \nu }$ by 
\begin{eqnarray}
E_{\mu \nu } [ \sqrt{ - g } \mathcal{F} ]  &:=&  \frac{ 1 }{ \sqrt{ - g } } \D _{ g^{\mu \nu } }  \sqrt{ - g } \mathcal{F} \\
\bar E_{\mu \nu } [ \sqrt{ - g } \mathcal{F} ]  &:=&  \frac{ 1 }{ \sqrt{ - g } } \bD _{ g^{\mu \nu } }  \sqrt{ - g } \mathcal{F} \\
	&=&  E _{ \mu \nu } [ \sqrt{ - g } \mathcal{F} ]  -  \frac{ 1 }{ 2 \sqrt{ - g } } u_{ \mu } u_{ \nu } u^{ \rho } \D _{ u^{ \rho } } \sqrt{ - g } \mathcal{F}  \\
F_{\mu \nu }  &:=&  \del_{\mu } A_{\nu }  -  \del_{\nu } A_{\mu } \\
	H _{ \mu \nu \rho }  &:=&  \del_{ \rho } B_{ \mu \nu }  +  \del_{ \nu } B_{ \rho \mu }  +  \del_{ \mu } B_{ \nu \rho }  ,  
\end{eqnarray}
and Using the relation for the covariant derivative of a symmetric tensor $S_{ \mu \nu }$ 
\begin{eqnarray}
\sqrt{ - g } \nabla _{ \nu } S^{ \nu } _{ \mu }  =  \del_{ \nu } \Big( \sqrt{ - g } S^{ \nu } _{ \mu } \Big)  +  \frac{ \sqrt{ - g } }{ 2 } \big( \del_{ \mu } g^{ \rho \sig } \big) S_{ \rho \sig } , 
\end{eqnarray}
it is found that the identity (\ref{id1}) is equivalent to 
\begin{eqnarray}
&&  2 \sqrt{ - g } \nabla _{\nu } \bar E^{\nu } _{\mu } [ \sqrt{ - g } \mathcal{F} ]   \nonumber \\
	&\equiv&  2 B_{ \mu \rho } \del_{ \nu } \Big( \D_{ \nu \rho } \sqrt{ - g } \mathcal{F} \Big)  -  H_{ \mu \nu \rho } \D_{ \nu \rho } \sqrt{ - g } \mathcal{F}  +  A_{\mu } \del_{\nu } \Big( \D _{ A_{\nu } }  \sqrt{ - g } \mathcal{F} \Big)  -  F_{\mu \nu } \D _{ A_{\nu } }  \sqrt{ - g } \mathcal{F}  \nonumber \\
	&&  -  \del_{ \nu } \Big[  u^{ \nu }  \big( u_{ \mu } u^{ \rho }  +  \delta ^{ \rho } _{ \mu }  \big)  \D _{ u^{ \rho } } \sqrt{ - g } \mathcal{F}  \Big]  -  \bigg[  \del_{ \mu } u^{ \rho }  -  \frac{ 1 }{ 2 } \big( \del_{ \mu } g_{ \alpha \beta } \big) u^{ \alpha } u^{ \beta } u^{ \rho }  \bigg]  \D _{ u^{ \rho } } \sqrt{ - g } \mathcal{F}   \nonumber \\
	&&  -  ( \del_{\mu } \varphi ) \D _{ \varphi }  \sqrt{ - g } \mathcal{F}  -  ( \del_{\mu } \eta ) \D _{ \eta }  \sqrt{ - g } \mathcal{F}  , \nonumber \\  \label{id1arrange}
\end{eqnarray}

Substituting the identity (\ref{id1}) into Eq. (\ref{deltaIntegral2a}), 
\begin{eqnarray}
0  &=&  \int _{ \mathcal{M} } \d ^n x \, \del_{\mu } \bigg[  \Big( 2  g^{\mu \beta } \D _{g^{\nu \beta } } \sqrt{ - g } \mathcal{F}  -  2 B_{ \nu \beta } \D_{ B_{ \mu \beta } } \sqrt{ - g } \mathcal{F}  -  A_{\nu } \D_{A_{\mu } }  \sqrt{ - g } \mathcal{F}  \nonumber \\
	&&  +  u^{ \mu } \D_{ u^{ \nu } } \sqrt{ - g } \mathcal{F}  -  \sqrt{ - g } T ^{\mu } _{\ \ \nu }  [ \sqrt{ - g } \mathcal{F} ]   \Big) \td x^{\nu }  +  \sum_{k = 0} ^{\infty }  \D _{\del_{\mu } \del_{\mu_1 } \cdots \del_{\mu_k } f }  \sqrt{ - g } \mathcal{F}  \cdot  \td \del_{\mu_1 } \cdots \del_{\mu_k } f  \bigg]  \nonumber \\
	\label{deltaIntegral3a} \\
	&=&  \int _{ \mathcal{M} } \d ^n x \, \bigg[  \del_{ \mu } \Big( 2  g^{\mu \beta } \D _{g^{\nu \beta } }  \sqrt{ - g } \mathcal{F}  -  2 B_{ \nu \beta } \D_{ B_{ \mu \beta } } \sqrt{ - g } \mathcal{F}  -  A_{\nu } \D _{A_{\mu } }  \sqrt{ - g } \mathcal{F}  \nonumber \\
	&&  +  u^{ \mu } \D_{ u^{ \nu } } \sqrt{ - g } \mathcal{F}  -  \sqrt{ - g } T ^{\mu } _{\ \ \nu }  [ \sqrt{ - g } \mathcal{F} ]   \Big) \td x ^{ \nu } \nonumber \\
		&&  +  \Big( 2  g^{\mu \beta } \D _{g^{\nu \beta } }  \sqrt{ - g } \mathcal{F}  -  2 B_{ \nu \beta } \D_{ \mu \beta } \sqrt{ - g } \mathcal{F}  - A_{\nu } \D _{A_{\mu } }  \sqrt{ - g } \mathcal{F}  \nonumber \\
		&&  +  u^{ \mu } \D_{ u^{ \nu } } \sqrt{ - g } \mathcal{F}  -  \sqrt{ - g } T ^{\mu } _{\ \ \nu }  [ \sqrt{ - g } \mathcal{F} ]   \Big) \del_{ \mu } \td x^{ \nu } \nonumber \\
		&&  +  \del_{ \mu } \Big( \sum_{k = 0} ^{\infty }  \D _{\del_{\mu } \del_{\mu_1 } \cdots \del_{\mu_k } f }  \sqrt{ - g } \mathcal{F}  \cdot  \td \del_{\mu_1 } \cdots \del_{\mu_k } f  \Big) \bigg] . \label{deltaIntegral3b}
\end{eqnarray}
Consider coordinate transformation such that $\del_{ \nu_1 } \td x^{ \mu } = \del_{ \nu_1 } \del_{ \nu_2 } \td x^{ \mu } = \cdots = 0$ on $\mathcal{M}$. Then, since $\td \del_{\mu_1 } \cdots \del_{\mu_k } f , \ \ \ k = 0,\ 1, \ 2, \ \cdots $ do not contain $\td x^{ \nu }$, the second and third line of (\ref{deltaIntegral3b}) vanish. Therefore, the coefficient of $\td x^{ \nu } $ is identically $0$, i.e., 
\begin{eqnarray}
	&&  \del_{\nu } \Big( 2 g^{\nu \beta } \D _{ g^{\mu \beta } }  \sqrt{ - g } \mathcal{F}  -  2 B_{ \mu \beta } \D_{ \nu \beta } \sqrt{ - g } \mathcal{F}  -  A_{\mu } \D _{ A_{ \nu } }  \sqrt{ - g } \mathcal{F}  \nonumber \\
	&&  +  u^{ \mu } \D_{ u^{ \nu } } \sqrt{ - g } \mathcal{F}  -  \sqrt{ - g } T^{\mu } _{\nu } [ \sqrt{ - g } \mathcal{F}]  \Big)  \equiv  0 .   \label{id2}
\end{eqnarray}

Substituting identity (\ref{id2}) into Eq. (\ref{deltaIntegral3b}), 
\begin{eqnarray}
0  &=&   \int _{ \mathcal{M} } \d ^n x \, \bigg[  \Big( 2  g^{\mu \beta } \D _{g^{\nu \beta } }  \sqrt{ - g } \mathcal{F}  -  2 B_{ \nu \beta } \D_{ B_{ \mu \beta } } \sqrt{  - g } \mathcal{F}  -  A_{\nu } \D _{A_{\mu } }  \sqrt{ - g } \mathcal{F}  \nonumber \\
	&&  +  u^{ \mu } \D_{ u^{ \nu } } \sqrt{ - g } \mathcal{F}  -  \sqrt{ - g } T ^{\mu } _{\ \ \nu }  [ \sqrt{ - g } \mathcal{F} ]   \Big) \del_{ \mu } \td x^{ \nu }   \nonumber \\
	&&  +  \del_{ \mu } \Big( \sum_{k = 0} ^{\infty }  \D _{\del_{\mu } \del_{\mu_1 } \cdots \del_{\mu_k } f }  \sqrt{ - g } \mathcal{F}  \cdot  \td \del_{\mu_1 } \cdots \del_{\mu_k } f  \Big) \bigg]  \label{deltaIntegral4a} \\
	&=&   \int _{ \mathcal{M} } \d ^n x \, \bigg[  \Big( 2  g^{\mu \beta } \D _{g^{\nu \beta } }  \sqrt{ - g } \mathcal{F}  - 2 B_{ \nu \beta } \D_{ B_{ \mu \beta } } \sqrt{ - g } \mathcal{F}  -  A_{\nu } \D _{A_{\mu } }  \sqrt{ - g } \mathcal{F}  \nonumber \\
	&&  +  u^{ \mu } \D_{ u^{ \nu } } \sqrt{ - g } \mathcal{F}  -  \sqrt{ - g } T ^{\mu } _{\ \ \nu }  [ \sqrt{ - g } \mathcal{F} ]   \Big) \del_{ \mu } \td x^{ \nu } \nonumber \\
		&&  + \del_{ \rho } \Big(  \Omega^{ \rho \mu } _{\ \ \ \nu } [ \sqrt{ - g } \mathcal{F} ]  \del_{ \mu } \td x^{ \nu }  +  \Xi ^{ \rho \mu_1 \mu_2 } _{\ \ \ \ \ \ \  \nu } [ \sqrt{ - g } \mathcal{F} ] \del_{ \mu_1 } \del_{ \mu_2 } \td x^{ \nu }  +  \cdots \Big) \bigg] . \label{deltaIntegral4b}
\end{eqnarray}
From Eqs. (\ref{deltaphidels}) - (\ref{deltaBdels1}), the coefficients, $\Omega ^{\rho \mu } _{\ \ \  \nu } $, $\Xi ^{ \rho \mu_1 \mu_2 } _{\ \ \ \ \ \ \  \nu } $, $\cdots $, can be calculated. For example, $\Omega ^{\rho \mu } _{\ \ \  \nu } $ is written explicitly: 
\begin{eqnarray}
&& \Omega ^{ \rho \mu } _{ \ \ \  \nu } [ \sqrt{ - g } \mathcal{F} ] \nonumber \\
	&=&   \sum_{k = 0} ^{\infty } \Bigg\{  \Big[ \delta ^{\alpha } _{\nu } ( \del_{\mu_1 } \cdots \del_{\mu_k } g^{ \mu \beta } )  +  \delta ^{\beta } _{\nu } ( \del_{\mu_1 } \cdots \del_{\mu_k } g^{\alpha \mu } )  \nonumber \\
	&&  \ \ \ \ \ \ \ \ \ \  -  \sum_{ l = 1 } ^k ( \del_{\nu } \del_{ \mu_1 } \cdots \del_{ l - 1 } \del_{ l + 1 } \cdots \del_{ \mu_{ k } } g^{ \alpha \beta } ) \delta ^{ \mu } _{ \mu_l }  \Big]  \D _{\del_{\rho } \del_{\mu_1 } \cdots \del_{\mu_k } g^{\alpha \beta } }  \sqrt{ - g } \mathcal{F} \nonumber \\
	&&  + \Big[  - \delta _{\alpha } ^{\mu } ( \del_{\mu_1 } \cdots \del_{\mu_k } B_{ \nu \beta } )  -  \delta _{\beta } ^{ \mu } ( \del_{\mu_1 } \cdots \del_{\mu_k } B_{ \alpha \nu } )  \nonumber \\
	&&  \ \ \ \ \ \ \ \ \ \  -  \sum_{ l = 1 } ^k ( \del_{\nu } \del_{ \mu_1 } \cdots \del_{ l - 1 } \del_{ l + 1 } \cdots \del_{ \mu_{ k } } B_{ \alpha \beta } ) \delta ^{ \mu } _{ \mu_l }  \Big]  \D _{\del_{\rho } \del_{\mu_1 } \cdots \del_{\mu_k } B_{ \alpha \beta } }  \sqrt{ - g } \mathcal{F} \nonumber \\
	&& + \Big[ - \delta ^{\mu } _{\alpha } ( \del_{\mu_1 } \cdots \del_{\mu_k } A_{\nu } )  -   \sum_{ l = 1 } ^k ( \del_{\nu } \del_{ \mu_1 } \cdots \del_{ l - 1 } \del_{ l + 1 } \cdots \del_{ \mu_{ k } } A_{\alpha } ) \delta ^{ \mu } _{ \mu_l }  \Big]   \D _{\del_{\rho } \del_{\mu_1 } \cdots \del_{\mu_k } A_{\alpha } }  \sqrt{ - g } \mathcal{F} \nonumber \\
	&&  +  \Big[  \delta^{ \alpha }_{ \nu } ( \del_{ \mu_1 } \cdots \del_{ \mu_k } u^{ \mu } )  -\sum_{ l = 1 } ^k ( \del_{\nu } \del_{ \mu_1 } \cdots \del_{ l - 1 } \del_{ l + 1 } \cdots \del_{ \mu_{ k } } u^{\alpha } ) \delta ^{ \mu } _{ \mu_l }  \Big]  \D_{ \del_{ \rho } \del_{ \mu_1 } \cdots \del_{ \mu_k } u^{ \alpha } } \sqrt{ - g } \mathcal{F}  \nonumber \\
	&&  - \sum_{ l = 1 } ^k \big( \del_{ \nu } \del_{ \mu_1 } \cdots \del_{ l - 1 } \del_{ l + 1 } \cdots \del_{ \mu_{ k } } \varphi  \big) \delta ^{ \mu } _{ \mu_l }  \D _{\del_{\rho } \del_{\mu_1 } \cdots \del_{\mu_k } \varphi }  \sqrt{ - g } \mathcal{F}   \Bigg\} . \label{defOmega}
\end{eqnarray}
The invariance of Eq. (\ref{deltaIntegral4b}) under coordinate transformations such that $\del_{ \mu_1 } \del_{ \mu_2 } \td x^{ \nu } = \del_{ \mu_1 } \del_{ \mu_2 } \del_{ \mu_3 } \td x^{ \nu } = \cdots = 0$ and $\del_{ \mu_1 } \del_{ \mu_2 } \del_{ \mu_3 } \td x^{ \nu } = \del_{ \mu_1 } \del_{ \mu_2 } \del_{ \mu_3 } \del_{ \mu_4 } \td x^{ \nu } = \cdots = 0$ requires the following equations to hold, respectively: 
\begin{eqnarray}
&& 2  g^{ \mu \beta } \D _{ g^{ \beta \nu } } \sqrt{ - g } \mathcal{F}  -  2 B_{ \nu \beta } \D_{ B_{ \mu \beta } } \sqrt{ - g } \mathcal{F}  -  A_{\nu } \D _{ A_{\mu } }  \sqrt{ - g } \mathcal{F}  +  u^{ \mu } \D_{ u^{ \nu } } \sqrt{ - g } \mathcal{F}  \nonumber \\
	&&  -  \sqrt{ - g } T ^{\mu } _{\ \ \nu }  [\sqrt{ - g } \mathcal{F}]  + \del_{ \rho } \Omega ^{ \rho \mu } _{ \ \ \  \nu } [ \sqrt{ - g } \mathcal{F} ]  \equiv  0   \nonumber \\
&& \label{id3} \\ 
&& \frac{1}{2} \Big( \Omega ^{ \mu_1 \mu_2 } _{ \ \ \ \ \ \  \nu } [ \sqrt{ - g } \mathcal{F} ]  +  \Omega ^{ \mu_2 \mu_1 } _{ \ \ \ \ \ \  \nu } [ \sqrt{ - g } \mathcal{F} ]  \Big)  +  \del_{ \rho } \Xi ^{ \rho \mu_1 \mu_2 } _{ \ \ \ \ \ \ \  \nu } [ \sqrt{ - g } \mathcal{F} ]  \equiv  0 . \label{id4}
\end{eqnarray}
Combining identities (\ref{id2}) and (\ref{id3}), the quantity $\del_{ \mu } \del_{ \rho } \Omega ^{ \rho \mu } _{ \ \ \  \nu } [ \sqrt{ - g } \mathcal{F} ]$ identically vanishes: 
\begin{eqnarray}
\del_{ \mu } \del_{ \rho } \Omega ^{ \rho \mu } _{ \ \ \  \nu } [ \sqrt{ - g } \mathcal{F} ]  \equiv  0 . \label{consvO}
\end{eqnarray}
Therefore, the quantity $\del_{ \rho } \Omega ^{ \rho \mu } _{ \nu } [ \sqrt{ - g } \mathcal{F} ]$ can be regarded as an energy-momentum peudotensor\footnote{The quantity $\del_{ \rho } \Omega ^{ \rho \mu } _{ \ \ \  \nu } [ \sqrt{ - g } \mathcal{F} ]$ does not transform as a tensor by coordinate transformations. } including the gravitational field. 
\begin{eqnarray}
M_{\rm Noerther}  :=  - \del_{ \rho }  \Omega ^{ \rho 0 } _{ \ \ \  0 } [ \sqrt{ - g } \mathcal{F} ]  \\
( P_{\rm Noether} )_a  :=   - \del_{ \rho }  \Omega ^{ \rho 0 } _{ \ \ \  a } [ \sqrt{ - g } \mathcal{F} ]  
\end{eqnarray}
are conserved charges, which can be interpreted as energy and momentum of the fields including the gravity, respectively.

Let us deform the identity (\ref{id3}) into the form used in section \ref{MEPtoFE}. Contracting Eq. (\ref{id3}) with $u_{ \mu } u^{ \nu } $, 
\begin{eqnarray}
	&& 2 u^{ \beta } u^{ \nu } \D _{ g^{\beta \nu } }  \sqrt{ - g } \mathcal{F}  +  u_{ \nu } \D _{ u_{ \nu } } \sqrt{ - g } \mathcal{F}  \nonumber \\
	&&  -  2 u_{ \mu } u^{ \nu } B_{ \nu \beta } \D _{ B_{ \mu \beta } } \sqrt{ - g } \mathcal{F}  -  u^{ \nu } A_{ \nu } u_{ \mu } \D _{ A_{ \mu } } \sqrt{ - g } \mathcal{F}  -  \sqrt{ - g } \mathcal{F}   \nonumber \\
	&\equiv &
- u_{ \mu } u^{ \nu } \del_{\rho }  \Omega ^{\rho \mu } _{\ \ \ \nu } [ \sqrt{ - g } \mathcal{F} ]  +  \sum_{k= 0} ^{\infty } u_{ \mu } \D_{\del_{\mu } \del_{\mu_1 } \cdots  \del_{\mu_k } f }  \sqrt{ - g } \mathcal{F} \cdot  u^{ \nu } \del_{\nu } \del_{\mu_1 } \cdots \del_{\mu_k } f . \label{id3uu} \nonumber \\
\end{eqnarray}
Using the relation (\ref{defDbar}), the first two terms of Eq. (\ref{id3uu}) are combined and yield $\bD _{ g^{ \rho \nu } } $, i.e., 
\begin{eqnarray}
&& 2 u^{ \rho } u^{ \nu }  \bD _{ g^{\rho \nu } }  \sqrt{ - g } \mathcal{F}  -  2 u_{ \mu } u^{ \nu } B_{ \nu \beta } \D _{ B_{ \mu \beta } } \sqrt{ - g } \mathcal{F}  -  u^{ \nu } A_{ \nu } u_{ \mu } \D _{ A_{ \mu } } \sqrt{ - g } \mathcal{F}  -  \sqrt{ - g } \mathcal{F}   \nonumber \\
&\equiv &
- u_{ \mu } u^{ \nu } \del_{\rho }  \Omega ^{\rho \mu } _{\ \ \ \nu } [ \sqrt{ - g } \mathcal{F} ]  +  \sum_{k= 0} ^{\infty } u_{ \mu } \D _{\del_{\mu } \del_{\mu_1 } \cdots  \del_{\mu_k } f }  \sqrt{ - g } \mathcal{F} \cdot  u^{ \nu } \del_{\nu } \del_{\mu_1 } \cdots \del_{\mu_k } f . \nonumber \\  \label{id3bar}
\end{eqnarray}

\subsection{Identities from the $U( 1 )$ gauge transformation} \label{gauge}
Next, consider the transformation, 
\begin{eqnarray}
\varphi  ^{ \prime } ( x )  &=&  e ^{ i e \Lam } \varphi ( x )  \label{gaugephi} \\
A_{ \mu } ^{ \prime } ( x )  &=&  A_{ \mu } ( x )  +  \del_{ \mu } \Lam ( x ) , \label{gaugeA}
\end{eqnarray}
where $\Lam ( x )$ is an arbitrary function, and $e$ is the electric charge. This is called $U(1)$ gauge transformation. In this subsection, let us derive the identities that any $U(1)$ gauge symmetric function $\mathcal{L}$ satisfies. 

The infinitesimal transformation of $U(1)$ gauge transformation is 
\begin{eqnarray}
& \td x  =  0  \\
& \td \varphi  =  i e  \varphi \delta \Lam \\
& \td A_{\mu }  =  \del_{\mu }  \delta \Lam \\
& \td B_{ \mu \nu }  =  0 \\
& \td g^{\mu \nu }  =  0 . 
\end{eqnarray}
Note that since the gauge transformation does not contain the coordinate transformation, the variations $\t \delta$ and $\delta$ are equivalent: 
\[
\t \delta  =  \delta . 
\]
The variations of the derivatives of $\varphi$ and $A_{ \mu }$ are 
\begin{eqnarray}
\td \del_{\mu_1 } \cdots \del_{\mu_k } \varphi  &=&  i e \sum_{j = 0} ^k 
	\bigg(
    \begin{array}{c}
      k \\
      j 
    \end{array}
	\bigg) 
( \del_{\mu_{j + 1} } \cdots \del_{\mu_k } \varphi  ) \del_{\mu_1 } \cdots \del_{\mu_j } \delta \Lam \label{gaugephidels} \\
\td \del_{\mu_1 } \cdots \del_{\mu_k } A_{\mu }  &=&  \del_{\mu_1 } \cdots \del_{\mu_k } \del_{\mu } \delta \Lam . \label{gaugeAdels}
\end{eqnarray}
Substituting Eqs. (\ref{gaugephidels}), (\ref{gaugeAdels}), and $\td x^{ \mu } = 0$ into Eq. (\ref{deltaIntegral}), 
\begin{eqnarray}
0  &=&  \int _{\mathcal{M} } \d ^n x  \bigg[  \D _f \sqrt{ - g } \mathcal{F}  \cdot  \t \delta f  +  \del_{ \mu } \bigg( \sum_{ k = 0 } ^{ \infty } \D _{ \del_{ \mu } \del_{ \mu_1 } \cdots \del_{ \mu_k } f } \sqrt{ - g } \mathcal{F}  \cdot \t \delta \del_{ \mu_1 } \cdots \del_{ \mu_k } f  \bigg)   \bigg]  \nonumber \\
\ \\
	&=&   \int _{ \mathcal{M} } \d ^n x \bigg[ \D _{ A_{ \mu } } \sqrt{ - g } \mathcal{F}  \cdot  \del_{ \mu } \delta \Lam  +  i e \varphi \D _{ \varphi } \sqrt{ - g } \mathcal{F}  \cdot \delta \Lam   \nonumber \\
	&&  +  \del_{ \mu } \bigg(  \sum_{ k = 0 } ^{ \infty } \D _{ \del_{ \mu } \del_{ \mu_1 } \cdots \del_{ \mu_k } f }  \sqrt{ - g } \mathcal{F}  \cdot  \t \delta \del_{ \mu_1 } \cdots \del_{ \mu_k } f     \bigg)  \bigg]   \\
	&=&  \int _{ \mathcal{M} } \d ^n x \bigg\{ \bigg[ - \del_{ \mu } \Big(  \D _{ A_{ \mu } } \sqrt{ - g } \mathcal{F}  \Big)   +  i e \varphi \D _{ \varphi } \sqrt{ - g } \mathcal{F}  \bigg] \cdot \delta \Lam  \nonumber \\
	&&  +  \del_{ \mu } \bigg( \D _{ A_{ \mu } } \sqrt{ - g } \mathcal{F} \cdot \delta \Lam  +   \sum_{ k = 0 } ^{ \infty } \D _{ \del_{ \mu } \del_{ \mu_1 } \cdots \del_{ \mu_k } f }  \sqrt{ - g } \mathcal{F}  \cdot  \t \delta \del_{ \mu_1 } \cdots \del_{ \mu_k } f   \bigg) \bigg\}  . \label{deltaIntegral5}
\end{eqnarray}
Since the gauge function $\Lam$ is arbitrary, this integral is invariant under the gauge transformation such that the variation of the gauge function $\delta \Lam$ is $0$ on $\del \mathcal{M}$. Therefore, the square bracket in Eq. (\ref{deltaIntegral5}) should be identically $0$, i.e., 
\begin{eqnarray}
\del_{ \mu } \Big(  \D _{ A_{ \mu } } \sqrt{ - g } \mathcal{F}  \Big)  -  i e \varphi \D _{ \varphi } \sqrt{ - g } \mathcal{F}  \equiv  0 . \label{id5}
\end{eqnarray}

Substituting the identity (\ref{id5}) into Eq. (\ref{deltaIntegral5}), 
\begin{eqnarray}
0  &=&  \int _{ \mathcal{M} } \d ^n x \del_{ \mu } \bigg( \D _{ A_{ \mu } } \sqrt{ - g } \mathcal{F} \cdot \delta \Lam  +   \sum_{ k = 0 } ^{ \infty } \D _{ \del_{ \mu } \del_{ \mu_1 } \cdots \del_{ \mu_k } f }  \sqrt{ - g } \mathcal{F}  \cdot  \t \delta \del_{ \mu_1 } \cdots \del_{ \mu_k } f   \bigg)  \nonumber  \\
\ \\
	&=&   \int _{ \mathcal{M} } \d ^n x \bigg[ \Big( \del_{ \mu } \D _{ A_{ \mu } } \sqrt{ - g } \mathcal{F}  \Big) \cdot \delta \Lam  +  \D _{ A_{ \mu } } \sqrt{ - g } \mathcal{F} \cdot \del_{ \mu } \delta \Lam  \nonumber \\
	&&  +  \del_{ \mu } \bigg(  \sum_{ k = 0 } ^{ \infty } \D _{ \del_{ \mu } \del_{ \mu_1 } \cdots \del_{ \mu_k } f }  \sqrt{ - g } \mathcal{F}  \cdot  \t \delta \del_{ \mu_1 } \cdots \del_{ \mu_k } f  \bigg)  \bigg]  \\
	&=&   \int _{ \mathcal{M} } \d ^n x  \bigg\{  \del_{ \mu } \bigg(  \D _{ A_{ \mu } } \sqrt{ - g } \mathcal{F}  +  i e \sum_{ k = 0 } ^{ \infty } ( \del_{ \mu_1 } \cdots \del_{ \mu_k } \varphi ) \D _{ \del_{ \mu } \del_{ \mu_1 } \cdots \del_{ \mu_k } \varphi } \sqrt{ - g } \mathcal{F}  \bigg) \cdot \delta \Lam  \nonumber \\
	&&  +  \bigg[ \D _{ A_{ \mu } } \sqrt{ - g } \mathcal{F}  +  \del_{ \nu } \D _{ \del_{ \nu } A_{ \mu } } \sqrt{ - g } \mathcal{F}   +   i e \sum_{ k = 0 } ^{ \infty } ( \del_{ \mu_1 } \cdots \del_{ \mu_k } \varphi ) \D _{ \del_{ \nu } \del_{ \mu_1 } \cdots \del_{ \mu_k } \varphi  } \sqrt{ - g } \mathcal{F}  \nonumber  \\
	&&  +  i e \del_{ \nu }  \bigg(  \sum_{ k = 0 } ^{ \infty } k ( \del_{ \mu_2 } \cdots \del_{ \mu_k } \varphi  )  \D _{ \del_{ \nu } \del_{ \mu } \del_{ \mu_2 } \cdots \del_{ \mu_k } \varphi } \sqrt{ - g } \mathcal{F}  \bigg)  \bigg] \cdot \del_{ \mu } \delta \Lam  \nonumber \\
	&&  +  \cdots  \bigg\}  . \label{deltaIntegral6}
\end{eqnarray}
The integral (\ref{deltaIntegral6}) is $0$ no matter what the variation of the gauge function $\delta \Lam$, the derivatives $\del_{ \mu } \Lam$, $\del_{ \mu_1 } \del_{ \mu_2 } \delta \Lam$, $\cdots$, and the domain of the integration $\mathcal{M}$. Therefore, the coefficients of $\delta \Lam$, $\del_{ \mu } \delta \Lam$, $\del_{ \mu_1 } \del_{ \mu_2 } \delta \Lam$, $\cdots$ should be identically $0$. For example, the first two are 
\begin{eqnarray}
\del_{ \mu } \bigg(  \D _{ A_{ \mu } } \sqrt{ - g } \mathcal{F}  +  i e \sum_{ k = 0 }^{ \infty } ( \del_{ \mu_1 } \cdots \del_{ \mu_k } \varphi )  \D _{ \del_{ \mu } \del_{ \mu_1 } \cdots \del_{ \mu_k } \varphi } \sqrt{ - g } \mathcal{F}  \bigg)   \equiv   0 
\end{eqnarray}
and 
\begin{eqnarray}
&&   \D _{ A_{ \mu } } \sqrt{ - g } \mathcal{F}  +  \del_{ \nu } \D _{ \del_{ \nu } A_{ \mu } } \sqrt{ - g } \mathcal{F}   +   i e \sum_{ k = 0 } ^{ \infty } ( \del_{ \mu_1 } \cdots \del_{ \mu_k } \varphi ) \D _{ \del_{ \nu } \del_{ \mu_1 } \cdots \del_{ \mu_k } \varphi  } \sqrt{ - g } \mathcal{F}  \nonumber  \\
&&  +  i e \del_{ \nu }  \bigg(  \sum_{ k = 0 } ^{ \infty } k ( \del_{ \mu_2 } \cdots \del_{ \mu_k } \varphi  )  \D _{ \del_{ \nu } \del_{ \mu } \del_{ \mu_2 } \cdots \del_{ \mu_k } \varphi } \sqrt{ - g } \mathcal{F}  \bigg)   \equiv   0 . 
\end{eqnarray}

\bibliographystyle{h-elsevier}
\bibliography{Literature}

\end{document}